\documentclass[twocolumn]{aastex62}
\usepackage[titletoc,title]{appendix}
\usepackage{longtable}
\usepackage{graphicx}
\usepackage{amsmath,amssymb}
\usepackage{color}
\usepackage{units}
\usepackage{epstopdf}
\usepackage{hyperref}
\usepackage{multirow}
\usepackage{float}
\usepackage{tikz}
\usetikzlibrary{shapes,arrows}

\usepackage{listings}
\usepackage{color}

\definecolor{dkgreen}{rgb}{0,0.6,0}
\definecolor{gray}{rgb}{0.5,0.5,0.5}
\definecolor{mauve}{rgb}{0.58,0,0.82}

\newcommand{\beq}{\begin{equation}}
\newcommand{\eeq}{\end{equation}}
\newcommand{\bdm}{\begin{displaymath}}
\newcommand{\edm}{\end{displaymath}}

\newcommand{\astrorapid}{\texttt{astrorapid}}

\definecolor{Gray}{gray}{0.9}
\definecolor{orange}{rgb}{0.9,0.5,0}

\graphicspath{{./plots/}}
\begin{document}

\title{Using machine learning for transient classification in searches for gravitational-wave counterparts}

\author{Cosmin Stachie}
\affil{Artemis, Universit\'e C\^ote d'Azur, Observatoire C\^ote d'Azur, CNRS, CS 34229, F-06304 Nice Cedex 4, France}

\author{Michael W. Coughlin}
\affil{Division of Physics, Math, and Astronomy, California Institute of Technology, Pasadena, CA 91125, USA}
\affil{School of Physics and Astronomy, University of Minnesota, Minneapolis, Minnesota 55455, USA}

\author{Nelson Christensen}
\affil{Artemis, Universit\'e C\^ote d'Azur, Observatoire C\^ote d'Azur, CNRS, CS 34229, F-06304 Nice Cedex 4, France}

\author{Daniel Muthukrishna}
\affil{Institute of Astronomy, University of Cambridge, Madingley Road, Cambridge CB3 0HA, UK}

\begin{abstract}
The large sky localization regions offered by the gravitational-wave
interferometers require efficient follow-up of the many counterpart
candidates identified by the wide field-of-view telescopes. Given the
restricted telescope time, the creation of prioritized lists of the
many identified candidates becomes mandatory. Towards this end, we use \text{\astrorapid}, a multi-band photometric lightcurve classifier, to differentiate between kilonovae, supernovae, and other possible transients. We demonstrate our method on the photometric observations of real events.
In addtion, the classification performance is tested on simulated lightcurves, both ideally and realistically sampled. We show that after only a few days of observations of an astronomical object, it is possible to rule out candidates as supernovae and other known transients.
\end{abstract}

\keywords{gravitational waves}

\section{Introduction}
\label{sec:Intro}

The first detection of a binary neutron star system GW170817 \citep{AbEA2017b} by the gravitational-wave (GW) detectors Advanced LIGO and Advanced Virgo was accompanied by the detection of both a short gamma-ray burst (SGRB) by {\it Fermi} Gamma-Ray Burst Monitor (GBM) \citep{AbEA2017c,AbEA2017d,AbEA2017e} and a kilonova by many other facilities \citep{CoFo2017,SmCh2017,AbEA2017f}.
This \emph{kilonova} is the ultra-violet/optical/infrared 
emission powered by the neutron-rich outflows undergoing the radioactive decay of r-process elements \citep{LaSc1974,LiPa1998,MeMa2010,KaMe2017}. 
The specifics of the lightcurves of kilonovae depend on the equation of state (EOS) of neutron stars and the mass ratio of the binary 
\citep{BaBa2013,PiNa2013,AbEA2017b,BaJu2017,DiUj2017,RaPe2018}.
In addition to this, there is synchrotron emission, which arises from a compact central engine launching a highly relativistic jet of electron/positron/baryon plasma \citep{WiRe1997,MeRe1998}. The internal dissipation of the jet's energy is responsible for the production of gamma rays and hard X-rays. The \emph{afterglow} phase, produced by interaction of the jet with the ambient material, consists of long lasting multi-wavelength emission in the X-ray, optical, and radio.
These three possible electromagnetic signatures of GW events, the kilonova, the SGRB, and the afterglow, have different characteristics. The kilonova is a short-lived isotropic emission in the visible and near infrared spectrum, the SGRB is a beamed flare of high energy X-rays and gamma-rays having a duration lower than 2s, while the afterglow is a long-standing multi-wavelength transient.
There have been a number of examples in the literature of using the photometry of both afterglows \citep{TrPi2018,AsCo2018} and kilonovae \citep{Coughlin_2017,SmCh2017,CoDi2018} to place constraints on the character of the progenitor systems.

The joint observations of these systems are interesting for a variety of reasons, including the study of SGRB beaming, energetics, and galactic environment \citep{MeBe2012}. In addition, the study of the kilonova lightcurves provides precious information about the nucleosynthesis of heavy elements in the Universe~\citep{Watson:2019xjv, Drout1570, Pian:2017gtc, Kasen:2017sxr} and the Hubble constant \citep{AbEA2017h,CoDi2019}. The first SGRB detected in association with a kilonova was 130603B \citep{Tanvir:2013pia}, providing support for the existence of an un-beamed electromagnetic signature to compact binary mergers. Also GRB 150101B has been reported as an off-axis jet associated to a blue kilonova by \citep{Troja:2018ybt}, based on its resemblance to GRB 170817A. However the detection of the kilonova transient represents a difficult task given the large sky localizations provided by both the $\gamma$-ray satellites, such as the {\it Fermi} GBM and GW interferometers. The localizations released by the GW detectors, in particular, can be large, spanning $\approx 100-10,000\,\textrm{deg}^2$ \citep{Rover2007a, Fair2009,Fair2011,Grover:2013,WeCh2010,SiAy2014,SiPr2014,BeMa2015,EsVi2015,CoLi2015,KlVe2016}. While GRB detections only have 2D sky localization information, the strain measurement in GW events allows also for the computation of a luminosity distance and therefore complete 3D skymap information is provided. 

The large sky localization regions require the use of wide-field survey telescopes to be covered. Observing instruments such as the Panoramic Survey Telescope and Rapid Response System (Pan-STARRS) \citep{MoKa2012}, Asteroid Terrestrial-impact Last Alert System (ATLAS) \citep{ToDe2018}, the Zwicky Transient Facility (ZTF) \citep{Bellm2018,Graham2018}, and in the near future BlackGEM \citep{10.1117/12.2232522} and the Large Synoptic Survey Telescope (LSST) \citep{Ivezic2014}, have the capabilities to observe these sky localizations. However, the main difficulty remains in the large number of  contaminant transients detected each night by these surveys. Among them, the identification of the counterpart represents a significant challenge. For this reason, an effective follow-up requires coordination between the wide FOV telescopes discovering transients and the telescopes that perform the follow-up and characterization of those transients. The transient characterization is typically done by smaller FOV telescopes performing both photometry and spectroscopy. Since the available telescope time at these telescopes is limited, it is essential to minimize the number of candidates that require observations, and be as efficient as possible with the classification. 

The challenging follow-up of electromagnetic transients has pushed the astronomical community to find new ways to optimize searches. An example is the formation of telescope networks \citep{Antier:2019pzz, Coughlin:2019xfb} which are generally built around the previously mentioned synoptic systems. Improved ways of tiling and observing these localizations using allocated time of telescopes are being employed based on the GW trigger candidate, telescope configuration, and possible electromagnetic counterparts \cite{10.1093/mnras/sty1066}. New proposals for the amelioration of the galaxy targeting strategy have been suggested, such as the prioritization of galaxies based on their stellar mass \citep{Ducoin:2019rdv}.

Techniques to optimize the follow-up of objects have been proposed in the literature. In \cite{Coughlin:2019xfb}, a method which combines the automated filtering and human vetting is used in order to reduce the number of initial candidates for S190425z \citep{2019GCN.24168....1L}, the first binary neutron star candidate from the third Advanced LIGO - Advanced Virgo observing run. During the automatic analysis, asteroids or near-Earth objects are removed as they did not appear in consecutive observations separated by a few tens of minutes. Objects very close ($<\,2$ arcsec) to point-like sources or having a historical detection prior to three days before the trigger are also automatically rejected. Finally, machine learning algorithms are used to identify image artifacts. Altogether, due to this automated filtering, it was possible to reduce the number of candidates from more than 300,000 to less than 300. Then, human vetting kept only those triggers which are in the localization, both in the 2D and distance, and which exhibited a rapid color evolution consistent with a kilonova. At the end of the entire analysis, fewer than 20 candidates remained. As another example, in \cite{Andreoni:2019qgh}, one can see the importance of having data not only after the GW trigger, but also prior to it. The DECam follow-up of the GW alert S190814bvs showed that it was difficult to rule out candidates due to the lack of recent pre-imaging history, resulting in significantly more candidates despite a smaller localization to cover. A proposal for improving the training set destined to a machine learning algorithm whose purpose is to identify supernovae photometric lightcurves is in \cite{Ishida:2018uqu}.

There are two kinds of objects that exhibit time variability: astrophysical objects whose signal last for a limited time and objects with periodic variation of the flux. Our knowledge about these electromagnetic events has dramatically improved in the last years. For some types, such as supernovae, there are models for their associated lightcurves, estimates of the occurence rate, as well as studies of their host-galaxy environment. Because of this, it is possible to simulate what a specific telescope will observe taking into account the instrument sensitivity and sky background. This was performed, for example, in the LSST \textit{PLAsTiCC} data challenge \citep{Kessler:2019qge}. In this study, they considered both extragalactic and galactic transients. Then, by means of the SuperNova ANAlysis software (SNANA) \citep{Kessler_2009}, a realistic set of lightcurves is generated illustrating what would be the LSST detections of transients of this type in the coming years. In this paper, we will use \text{\astrorapid} \citep{Muthukrishna:2019wgc}, a classifier tool based on machine learning to classify objects. It was trained on a set of lightcurves generated using SNANA and \textit{PLAsTiCC} in order to simulate a realistic set of events that would be observed by ZTF. \text{\astrorapid}  is designed to distinguish between transient templates.

We will evaluate the ability to use machine learning classifiers on early photometric lightcurves to support prioritization of transients for follow-up in GW and SGRB follow-up. While we will focus on kilonovae, the technique will be suitable for detection of afterglows.
We will describe the algorithm we use in Section~\ref{sec:algorithm}.
In section~\ref{sec:performance}, we describe the performance of the algorithms.
In section~\ref{sec:conclusions}, we offer concluding remarks and suggest directions for future research. The paper ends with Appendix ~\ref{sec:penalty factor} presenting the statistical justification for some classification criteria introduced in section~\ref{sec:algorithm}.

\section{Algorithm}
\label{sec:algorithm}

\begin{figure}[!htb]
    \centering
    \includegraphics[scale=0.3]{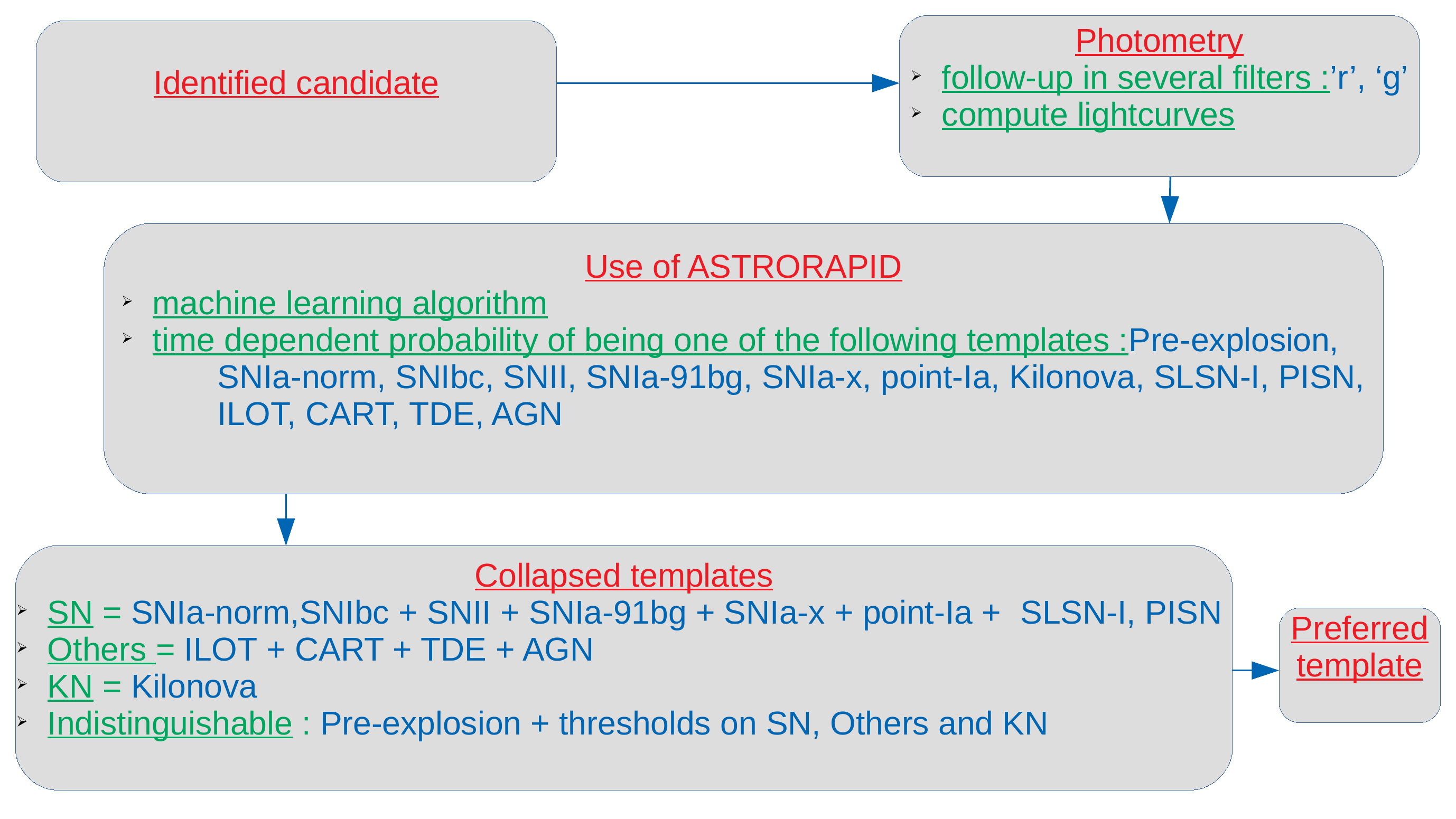}
    \caption{Flow chart illustrating the different steps made in the photometry analysis. The starting point is represented by the initial set of identified candidates, after which optical observations are carried out for each of these events. Then \text{\astrorapid} provides a time dependent probability distribution spread over fourteen possible candidate classes. The results provided by \text{\astrorapid} are handled in order to give weights and  discriminate between only four main classes: ``SN'', ``KN'', ``Others'' and ``Indistinguishable.'' Finally, a preferred class is declared.} 
    \label{Fig:flowchart}
\end{figure}

The idea of this analysis is to use multi-epoch photometry to identify interesting candidates. For the kilonova models being explored here, significant changes in magnitude are expected on time-scales of a single night. For this reason, telescope network photometry will determine which transients can feasibly be related to the event, and otherwise determine the background supernovae and other unrelated transients.
A flow chart showing up the method used in this study (and explained further in the following) to identify and characterize optical counterparts is shown in Figure~\ref{Fig:flowchart}. 

For this purpose, we use \text{\astrorapid}, which was developed to distinguish between fourteen different templates: ``Pre-explosion'' (a template introduced in order to distinguish the targeted flaring event from the moments preceding it~\citep{Muthukrishna:2019wgc}), ``SNIa-norm'' (a subtype of Type Ia Supernovae), ``SNIbc'' (a subtype of Core collapse Supernovae), ``SNII'' (a subtype of Core collapse Supernovae), ``SNIa-91bg'' (a subtype of Type Ia Supernovae), ``SNIa-x'' (a subtype of Type Ia Supernovae~\citep{Silverman_2012, Foley_2013})", ``point-Ia'' (hypothetical supernova type~\citep{shen2010thermonuclear}), ``Kilonova'' (electromagnetic counterpart to either binary neutron star (BNS) or neutron star - black hole (NS-BH) mergers~\citep{PhysRevLett.119.161101}), ``SLSN-I'' (Type I Super-luminous supernovae~\citep{article_quimby}), ``PISN'' (Pair-instability Supernovae~\cite{Ren_2012}), ``ILOT'' (Intermediate Luminosity Optical Transients~\citep{berger2009intermediate}), ``CART'' (Calcium-rich gap transients~\citep{lunnan2017two}), ``TDE'' (Tidal disruption Events~\citep{rees1988tidal}), and ``AGN'' (Active galactic nuclei). A description of each specific template can be found in \citealt{Kessler:2019qge}. At the same time the model libraries used in this analysis are publicly available\footnote{\url{https://doi.org/10.5281/zenodo.2612896}}. \citealt{Muthukrishna:2019wgc} describes the intrinsic luminosity, lightcurve shapes, and color evolution for the transient templates used in the training of \text{\astrorapid}. In addtion, Figure 2 of \citealt{Muthukrishna:2019wgc} displays example lightcurve shapes of the these templates.
The tool takes as input data the lightcurve of the transient, including the time of the exposure, apparent magnitude and associated error bar; the output consists of a time-dependent discrete probability distribution. It can take either a redshift (such as from a probable host galaxy) or not, and we will show results for both cases in the following.
This distribution provides the probability for each one of the 14 template types. The probability distribution changes with every new observation, and the more points the lightcurve contains, the more precise the identification.

While \text{\astrorapid} was designed to classify full lightcurves, the goal of this analysis is to determine, given a few observations, how to prioritize objects for follow-up to support kilonova identification.
To this end, we made a few modifications to the initial code. First of all, we collapse all fourteen  different templates into three main classes. Thus we consider the following classes: "SN" (which accounts for ``SNIa-norm,'' ``SNIbc,'' ``SNII,'' ``SNIa-91bg,'' ``SNIa-x,'' ``point-Ia,'' ``SLSN-I,'' ``PISN''), "Others" (accounting for ``ILOT,'' ``CART,'' ``TDE,'' and ``AGN'') and ``KN'' (which is simply the ``Kilonova'' template). The probability of ``Others'' and ``KN'' is simply the sum of the probabilities of their constituents. For the new probability for the ``SN'' class in particular, we found that we are required to penalize it more than the others (likely because it is made up of a majority of the classes). Therefore, for ``SN'' in particular, we use the sum of the probabilities of  its components multiplied by the factor $(1.- e^{- k^{\text{th}}_{\text{obs}} / \beta})$, where $k^{\text{th}}_{\text{obs}}$ stands for the $(k+1)^{\text{th}}$ observation. The choice of the factor $(1.- e^{- k^{\text{th}}_{\text{obs}} / \beta})$,  is based on the following considerations, and specific tests done with real data of transients  detected by ZTF as described in Appendix~\ref{sec:penalty factor}. Considering only the very first observations the probabilities of the initial \astrorapid \text{} templates are about equal, and thus the probability of the ``SN' class becomes large because it accounts for numerous initial templates, and needs to be reduced. We choose an exponential function to impose our knowledge that having only a few observations is uninformative (i.e. returning essentially the prior) and to leave unchanged the late observation predictions.  Hand-tuning for $\beta$ led to $\beta = 4 $, although it is worth acknowledging that this particular choice might not be optimal. Then, we also introduce a new class called ``Indistinguishable'', where we will say that the preferred event is ``Indistinguishable'' if none of the other classes (``SN,'' ``KN,'' and ``Others'') has a probability higher than $40\%$. The 40\% threshold was selected at the end of several trials. This value represents a trade off between two different behaviors. A higher threshold will favor too much the ``Indistinguishable'' class at least for the very first observations where there is not much information and the initial weight of the ``Pre-explosion'' template is already high. On the other hand, the consequence of a lower threshold will be to force the modified classifier to choose some class in \{"SN", "KN", "Others"\}, although there is not enough information for any inference; this is also an undesirable effect. In summary, in total there are 4 classes: ``KN,'' ``SN,'' ``Others'' and ``Indistinguishable.'' We also now introduce the idea of a ``preferred event'' after $k$ observations. We define the preferred event as X (here X stands for ``KN'', ``SN'' or ``Others'') if two conditions are fulfilled: (i) $Prob(\textnormal{X}) = \max(Prob(\textnormal{"SN"}), Prob(\textnormal{"KN"}), Prob(\textnormal{"Others"}))$ and (ii) $Prob(\textnormal{X}) > 40\%$. This will be convenient for classification later. More details about the motivation of both the penalty factor and the threshold can be found in Appendix ~\ref{sec:penalty factor}.

\section{Performance}
\label{sec:performance}

To demonstrate the utility of the method for transient prioritization and identification, we seek to show that two conditions hold. The first is that kilonova lightcurves should in general be identified in the ``KN'' class. The second is that for input lightcurves representing some other transient type, the analysis should not misidentify them as a ``KN.''
In the following, we will use both simulated lightcurves and real ZTF astrophysical transients from the public survey to assess these questions. For the injection sets in particular, we will create two sets of simulated SNe and KNe lightcurves. Each set has 1,000 lightcurves representing transients uniformly distributed in distance between 40\,Mpc and 3,000\,Mpc. The set of real ZTF events is formed by 2,291 lightcurves from the public data stream with an average of 29 observations per lightcurve. These events are mainly different types of supernovae, but also include transients like TDE, AGN, ILOT and CART. 

\subsection{Real transients observed in multiple filters}
\label{subsec: real transients}

\begin{figure*}[!htb]
    \centering
    \includegraphics[scale=0.158]{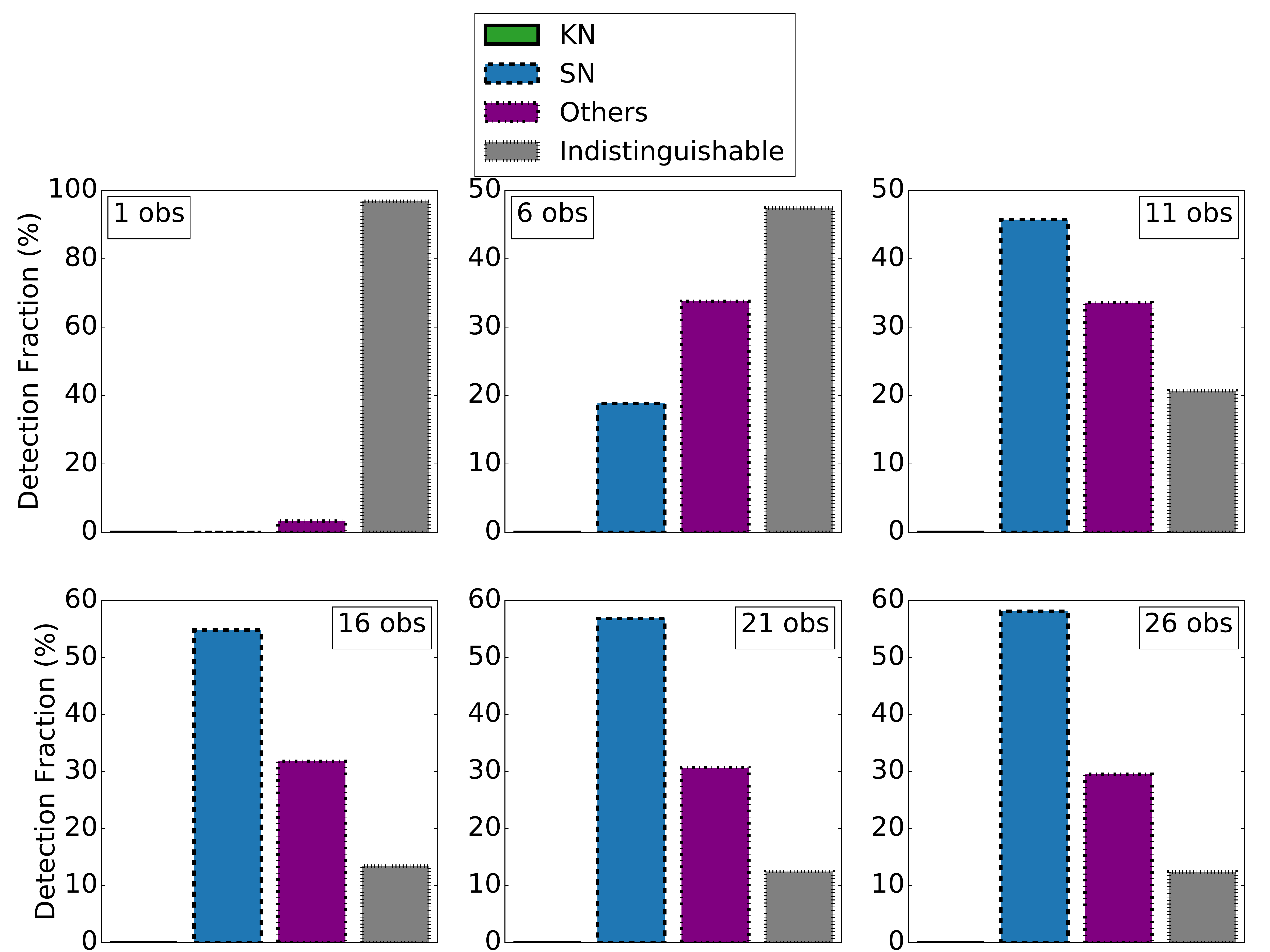}
    \includegraphics[scale=0.158]{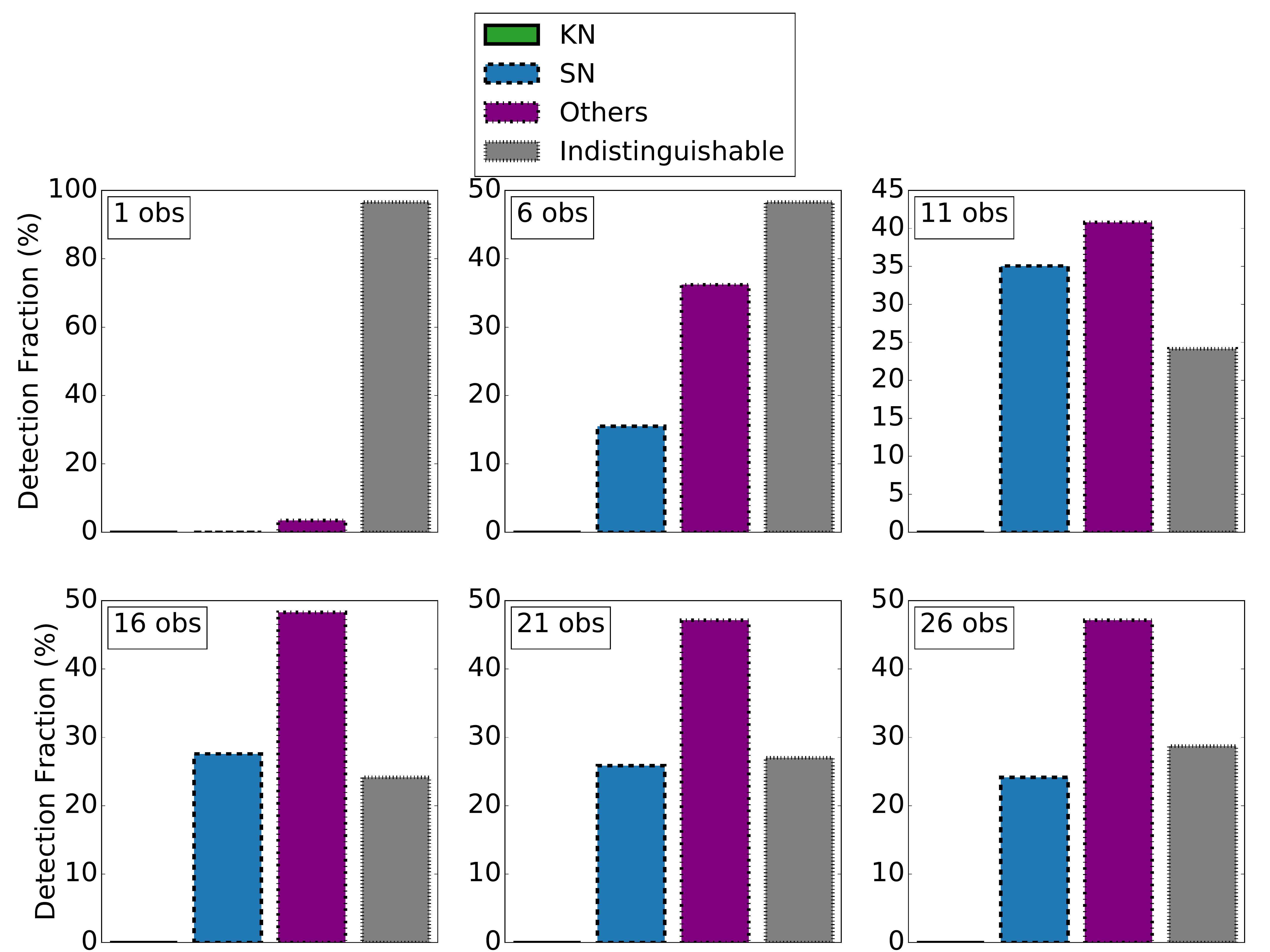}
    \caption{On the left are the histograms  of classifier favored events given different number of observations: from 1 (top-left corner) up to 26 (bottom-right corner). The input is represented by the set of 2,049 ZTF real sources identified as ``SN'' type. On the right is the same for the set of 174 ZTF real sources identified as the ``Others'' type.}
    \label{fig:ZTF-Real}
\end{figure*}

\begin{figure}[!htb]
    \centering
    \includegraphics[scale=0.18]{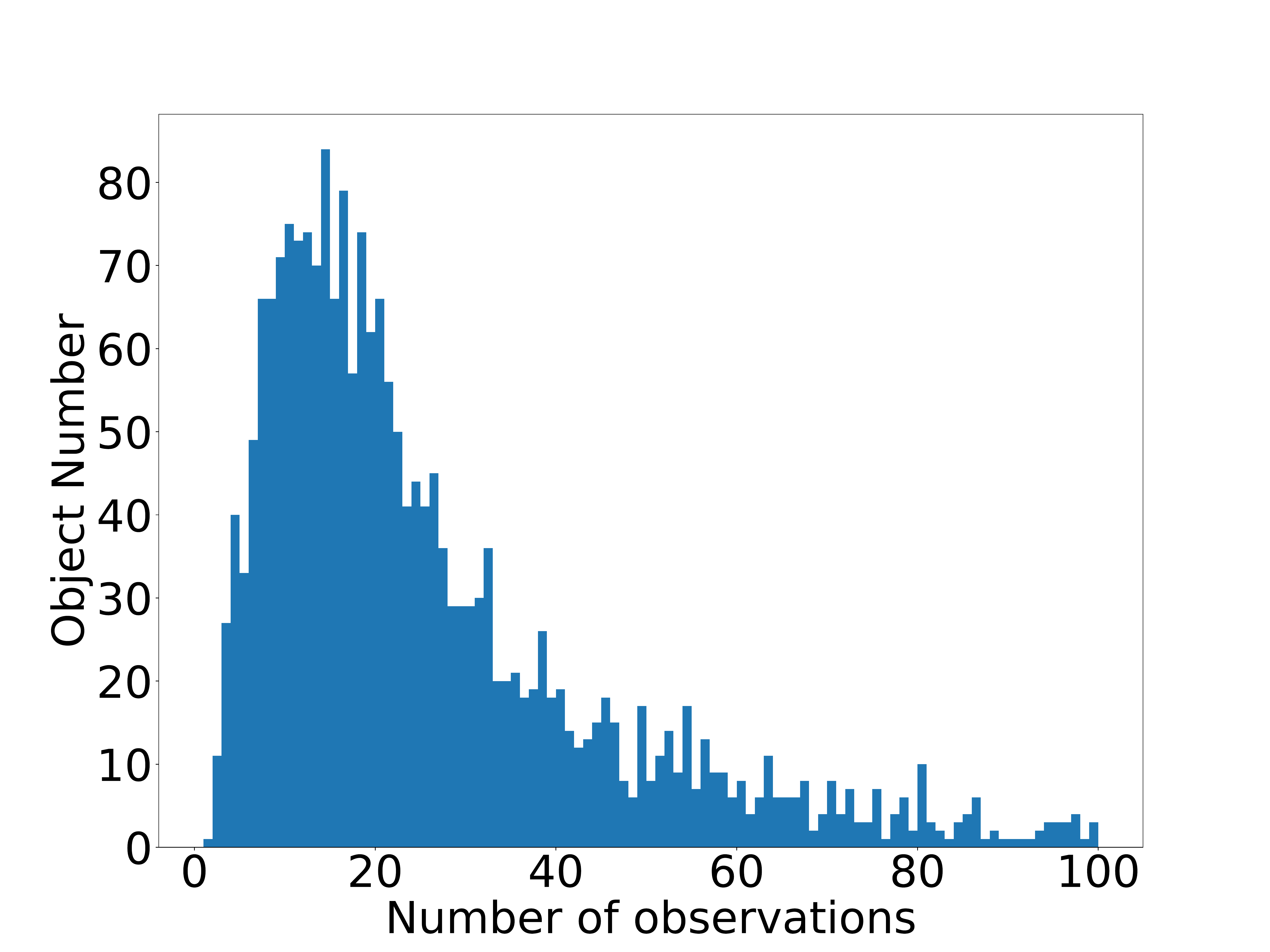}
    \caption{Histogram showing the observations number for the set of real ZTF objetcs. Both "SN" and "Others" types objects are included in this study.}
    \label{Fig:ZTF-nb_obs}
\end{figure}

The study of lightcurves observed during real survey is obviously necessary to assess the performances of our classifier in realistic situations. To this end, we use public ZTF lightcurves. We put these events in two categories: ``SN'' and ``Others.'' There are 2,049 ``SN'' type events (1450 ``SNIa,'' 110 ``SNIbc,'' 447 ``SNII,'' and 42 ``SLSN'') and 174 ``Others'' type events (152 ``AGN,'' 4 ``CART,'' 6 ``ILOT,'' and 12 ``TDE''). As input, we consider only the observational data from the $r$ and $g$-bands. Figure~\ref{fig:ZTF-Real} displays the results of the classifier after the first observation points in the case of ``SN'' and ``Others'' type real events. One can see that the classifier starts to correctly classify (efficiency higher than 40\%) after only 11 observations in the identification of ``SN'' and 16 in the ``Others'' case. It also exhibits that the classifier almost never misidentifies these real events as being kilonovae. This shows that kilonovae represent a fundamentally different part of the parameter space. As pointed out above, these real ZTF objects have an average of 29 observations per lightcurve. Figure~\ref{Fig:ZTF-nb_obs} illustrates the number of observations for the set of real ZTF objects. 
To take into account the unequal number of observations, we consider for a lightcurve possessing a total of $N$ observations, that the preferred event after $m$ observations with $m > N$, to be the same as the one found at the end of the $N$ (real) observations.

\subsection{Real transients observed in a single filter}

\begin{figure}[!htb]
    \centering
    \includegraphics[scale=0.18]{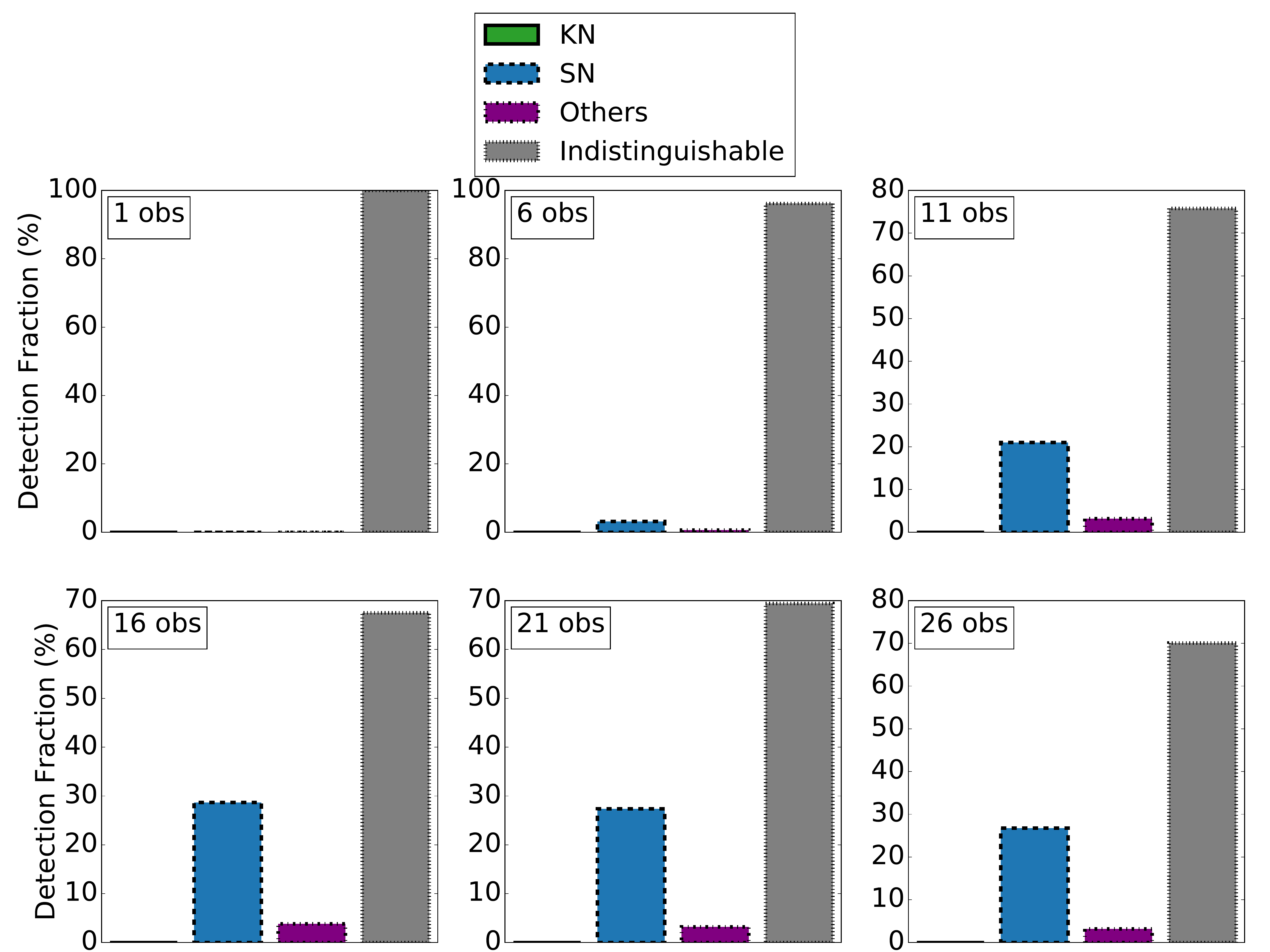}
    \caption{Preferred template fraction given different number of observations: from 1 (top-left corner) up to 26 (bottom-right corner). The input is represented by the set of 157 ZTF real sources observed in only one filter. Those events were finally all identified as being SN types.}
    \label{Fig:ZTF-SN_one_filter}
\end{figure}

Among the real supernovae transient lightcurves, there are 157 transients which have only single passband observations. As for any deep learning algorithm, the idea is that the more information there is, the better is the classification. 
Thus a classification results comparison between the two passbands ($r$ and/or $g$) case and only one passband (only $r$ or only $g$) case is worthwhile.
In Figure~\ref{Fig:ZTF-SN_one_filter}, we show the results of the follow-up of those single filter events. In this case, the success rate after 26 observations is much lower (around 10\%) compared to the case presented in the previous section (success rate around 60\%). This highlights how essential color information is in the classification. It is worth mentioning that simulations for which \text{\astrorapid} was trained on did not include single filter light curves, therefore as a future activity retraining \text{\astrorapid} to include both single and multi photometric bands will help to fix this issue.   

\subsection{Injection sets}
There are several reasons which motivate us to verify performance using simulated lightcurves. It was shown in the previous sections how well the classifier performs in identifying supernovae sampled at the ZTF cadence (an illustration of the ZTF cadence is available in Figure~\ref{fig:ZTF_cadence}). A question that arises is how much the background transient identification can be improved if we increase the observational rate, or equivalently, how much a dedicated ToO observation will ameliorate the results compared to the typical ZTF cadence. Likewise, we want to check how well the kilonova lightcurves are recovered by our classifier. Because of the very small set of real kilonovae detected to date, the choice of an injection set becomes important.

\begin{figure*}[!htb]
   \includegraphics[scale=0.17]{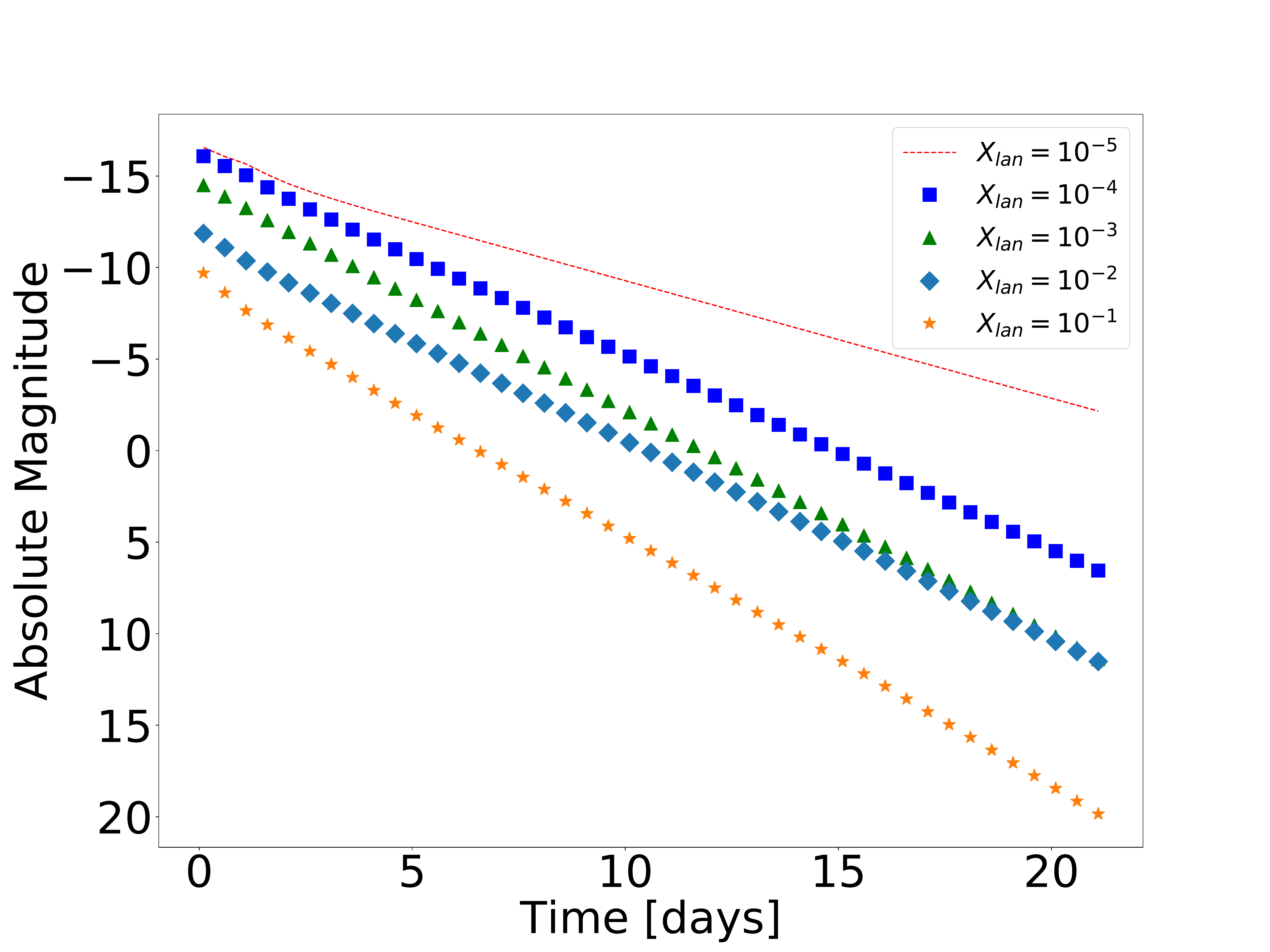}
   \includegraphics[scale=0.17]{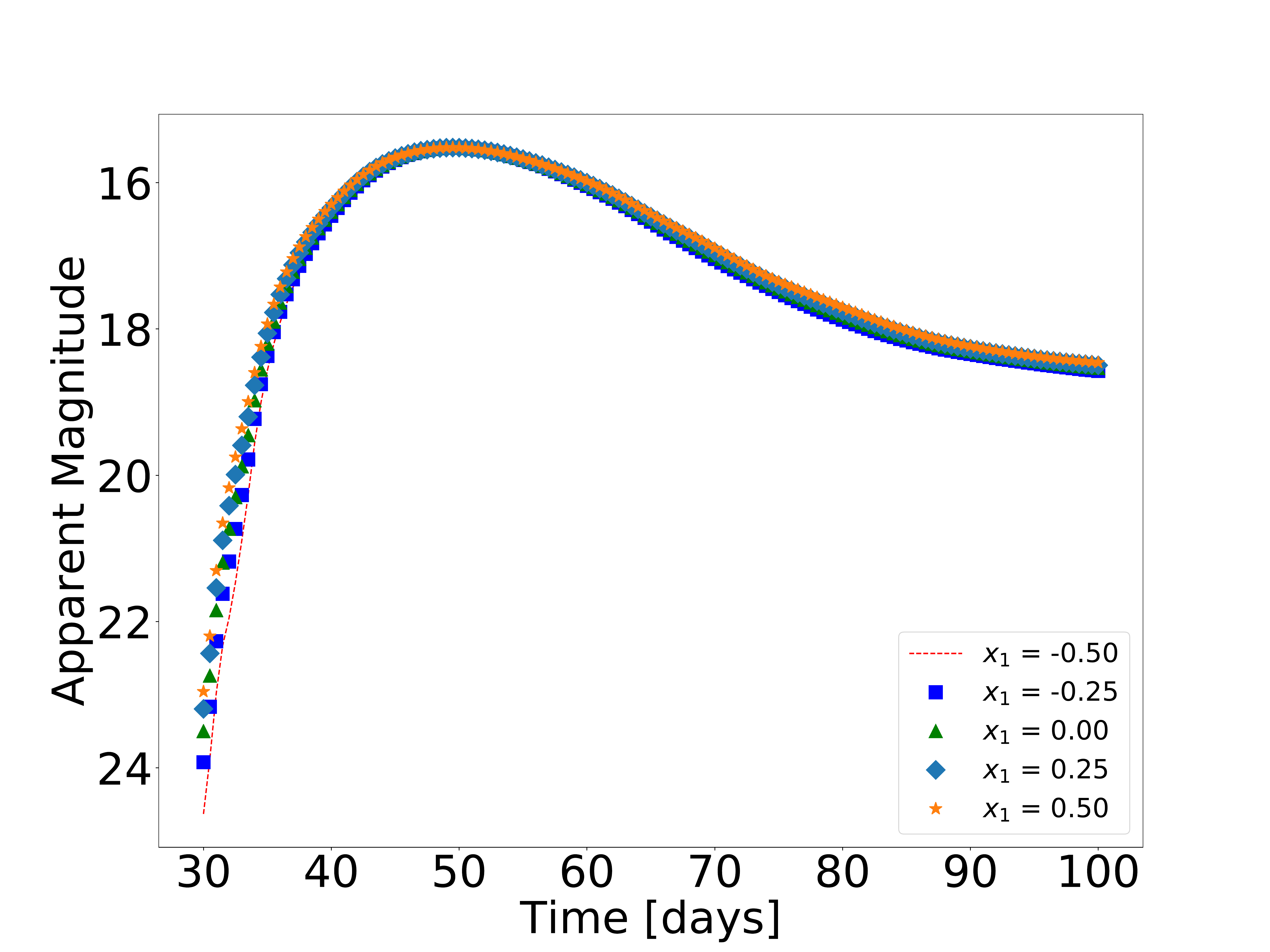}
   \caption{Left: Lightcurves representing the Absolute Magnitude as a function of time for five KN injections, for different input parameters. For all the lightcurves $M_{ej}$ = 0.05\(\textup{M}_\odot\) and $v_{ej}$ = 0.15c. The only parameter that varies is $X_{lan}$, whose value is shown on the legend. Right: Apparent Magnitude versus time for five SN injections. The lightcurves share parameters like cosmological redshift $z = 0.022$ and color index $c = 0.$ Here the shape parameter $x_1$ is different for each lightcurve and its value is shown in the legend. These are the simulation results for the $g$-band for both plots.}
   \label{fig:lightcurves}
\end{figure*}

As described above, we use simulations of both kilonova and supernova lightcurves generated by varying different parameters. The codebase used to generate kilonova lightcurves was described previously in~\cite{10.1093/mnrasl/slz133} and~\cite{ 2018MNRAS.480.3871C}. The variable physical parameters in the model are the ejecta mass ($M_{ej}$), the velocity of the ejecta ($v_{ej}$) and the lanthanide fraction ($X_{lan}$). Likewise, the prior range of the parameters are $M_{ej}$ $\in$ [0.01\(\textup{M}_\odot\), 0.1\(\textup{M}_\odot\)],  $v_{ej}$ $\in$ [0.01c, 0.3c] and $X_{lan}$ $\in$ [$10^{-5}$, $10^{-1}$], where \(\textup{M}_\odot\) is the solar mass and $c$
represents the speed of light. While $v_{ej}$ is sampled uniformly, the parameters $X_{lan}$ and $M_{ej}$ are log-uniformly distributed. An illustration of several such lightcurves is given on the right of Figure~\ref{fig:lightcurves}.
Type Ia supernovae lightcurves are generated by ``sncosmo,'' whose details are explained in~\cite{barbary2016sncosmo, Guy:2007dv}. In order to be as general as possible, parameters like parameter shape (hereafter $x_1$) and color index (hereafter $c$) were chosen to fill a broad space: $x_1 \in [-0.5, 0.5]$ and $c \in [-0.05, 0.05]$. The left of figure~\ref{fig:lightcurves} shows a few such lightcurves.

We want to imitate the ``background'' introduced by such lightcurves in searches for kilonovae. To this end, it is important to understand where the supernova observations are positioned relative to the peak of the lightcurve. If there are enough points before the peak, then one could expect a few misclassifications as ``KNe'' because a rising lightcurve is not characteristic for kilonovae. 
\begin{figure}[!htb]
    \centering
    \includegraphics[scale=0.18]{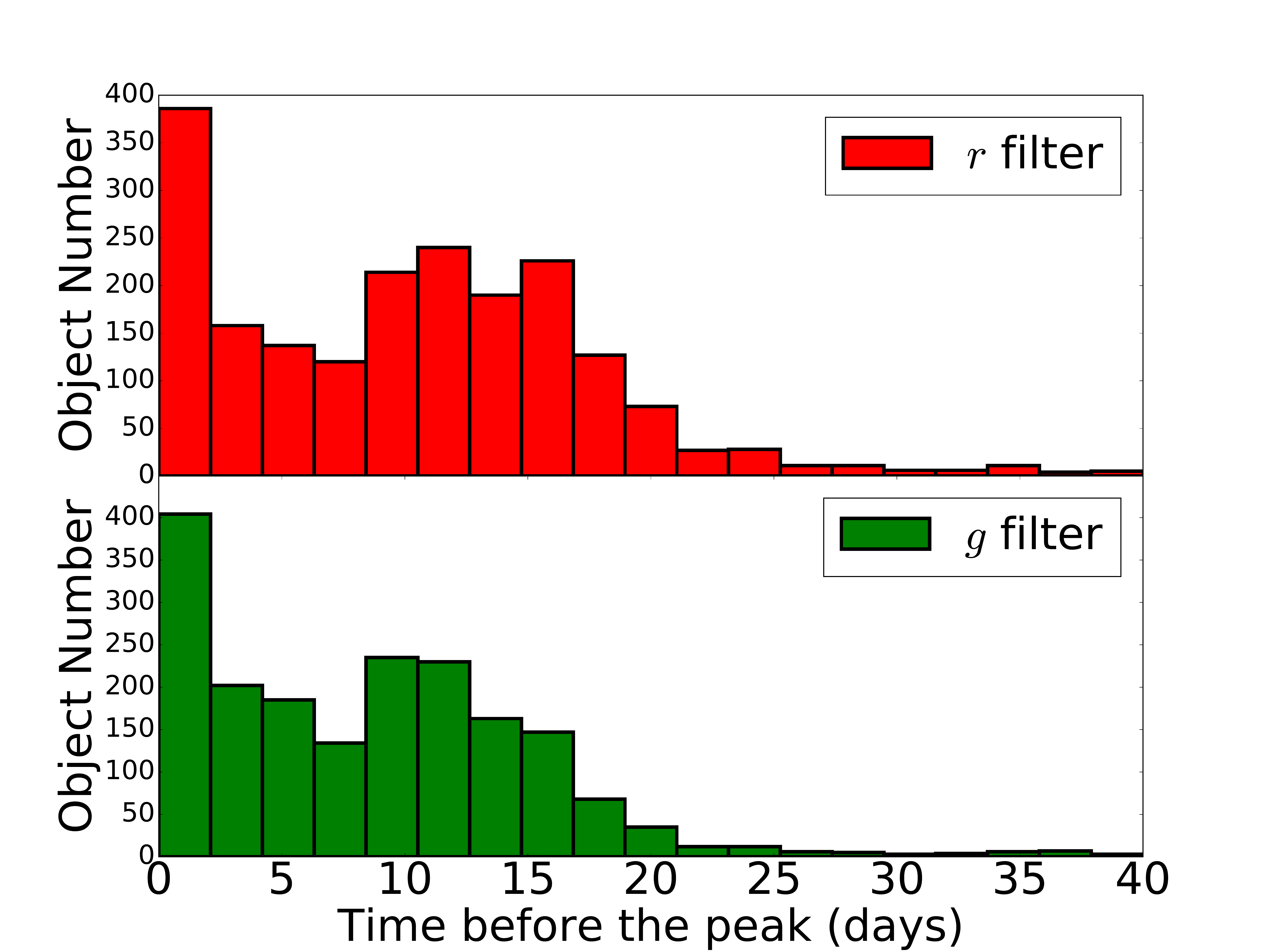}
    \caption{The amount of time before the peak of the lightcurve for the first observation. On the top (bottom) plot there is the histogram corresponding to the $r$ ($g$) passband. The input is represented by the set of real "SN" ZTF objects.}
    \label{fig:time_before_peak}
\end{figure}
In Figure~\ref{fig:time_before_peak}, we illustrate the time of the first observation with respect to the peak of the supernova lightcurve. It is worth mentioning that by first observation, we mean first detection of the supernova. The data set used in this study was the same set of real ``SN'' objects used in Section~\ref{subsec: real transients}. It is worth mentioning that we cannot be sure if the peak of the lightcurve coincides with the peak of the supernova, while in principle, it could be a local maxima. However, even a local maxima will appear as a ``rise,'' which is one of the main features we are investigating here. From Figure~\ref{fig:time_before_peak}, one sees that, for a non-negligible (more than 10\%) part of the objects, their first observation represents the peak of the lightcurve. We do not know if it coincides with the global peak, or where the global peak might actually be. According to these results, we choose the first observations of the supernovae to be uniformly distributed in the range [-7 days, + 30 days].

Looking at the distance distribution, we considered a uniformly distributed population of supernovae and kilonovae between 40\,Mpc and 300\,Mpc (chosen as the edge of the binary neutron star GW detection horizons). 
\begin{figure}[!htb]
    \centering
    \includegraphics[scale=0.18]{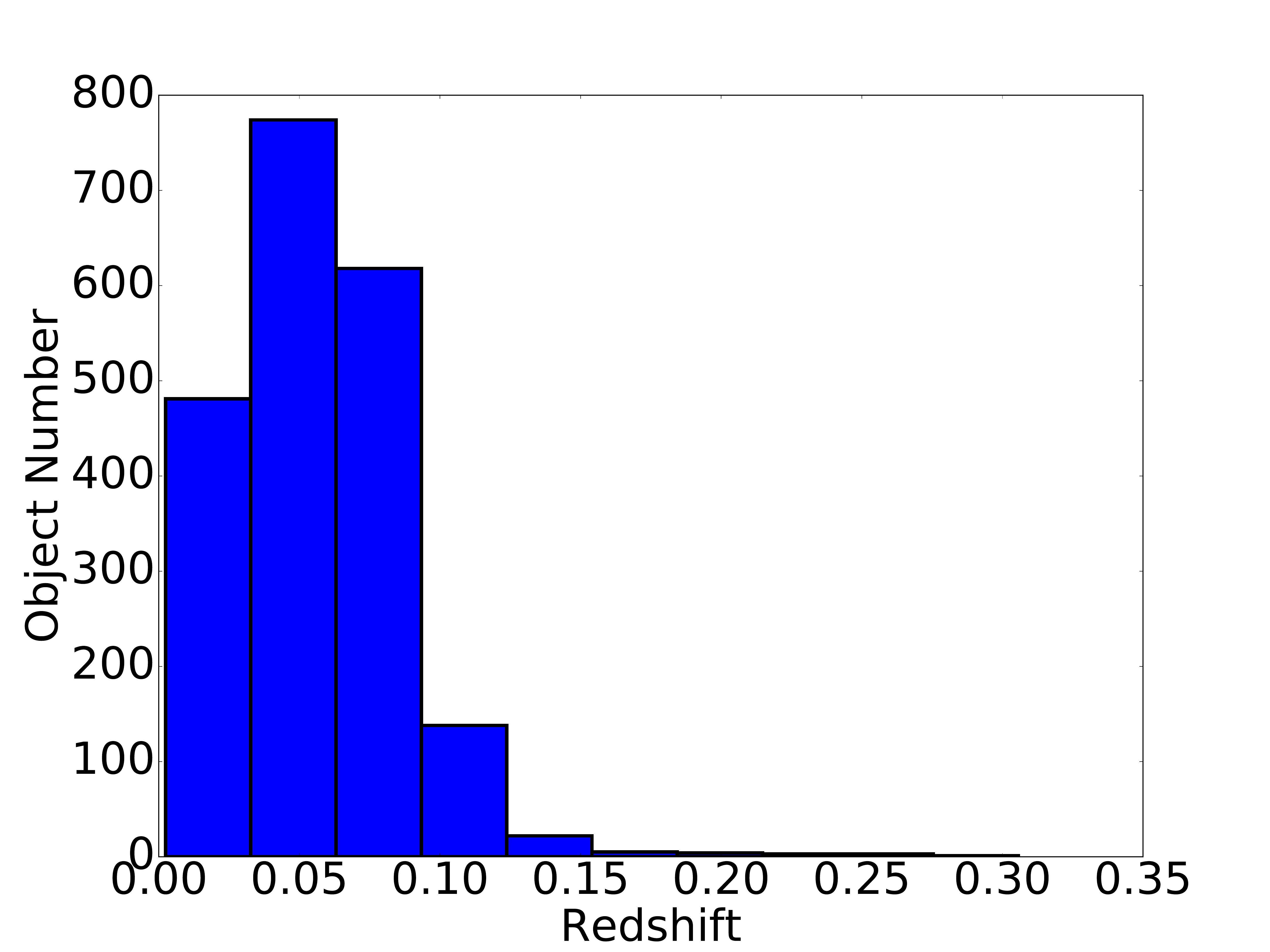}
    \caption{Redshift histogram for the real ZTF "SN" type lightcuves.}
    \label{fig:distribution_redshift}
\end{figure}
This choice is also motivated by the distribution of the cosmological redshifts in the case real ZTF ``SN'' objects; this is the same lightcurve set used in Section~\ref{subsec: real transients}. Indeed in Figure~\ref{fig:distribution_redshift}, one can see that this distribution possesses a peak around $z=0.05$, which corresponds to a luminosity distance of approximately 230\,Mpc.
Although a population uniformly distributed in volume is also possible, we injected uniformly in distance because we are predominantly interested in how the algorithm performs as a function of distance, and so how it performs spanning the bright to the faint end. A volume-limited set would be a more realistic distribution of kilonovae, and essential for any rates-related work.

In the case of kilonovae, we considered two sets, the first one ideally sampled (two observations per night per filter on average) and the second one realistically sampled (one observation per three nights per filter on average).
In the view of already treated real ZTF ``SN'' objects, we limit ourselves to only the case of ideally sampled supernovae injections. Magnitude uncertainties have been also taken into account for these sets of lightcurves. The error bars considered for this study are magnitude and filter dependent, and have similar values to those measured on the real ZTF objects.

We verify the performance of our method on the simulated lightcurves. In the case of kilonova injections, the output assesses how well the classifier recognizes a kilonova; the purpose of evaluating the output on supernovae injections is to quantify how often our tool misidentifies a transient as being a kilonova when it is not. In this regard, the supernova transients represent our ``background.'' The choice of supernovae as the background set is motivated by the dominance of this type of transient among the identified candidates. 

\subsubsection{Well-sampled Type Ia supernovae}
\begin{figure}[!htb]
    \centering
    \includegraphics[scale=0.18]{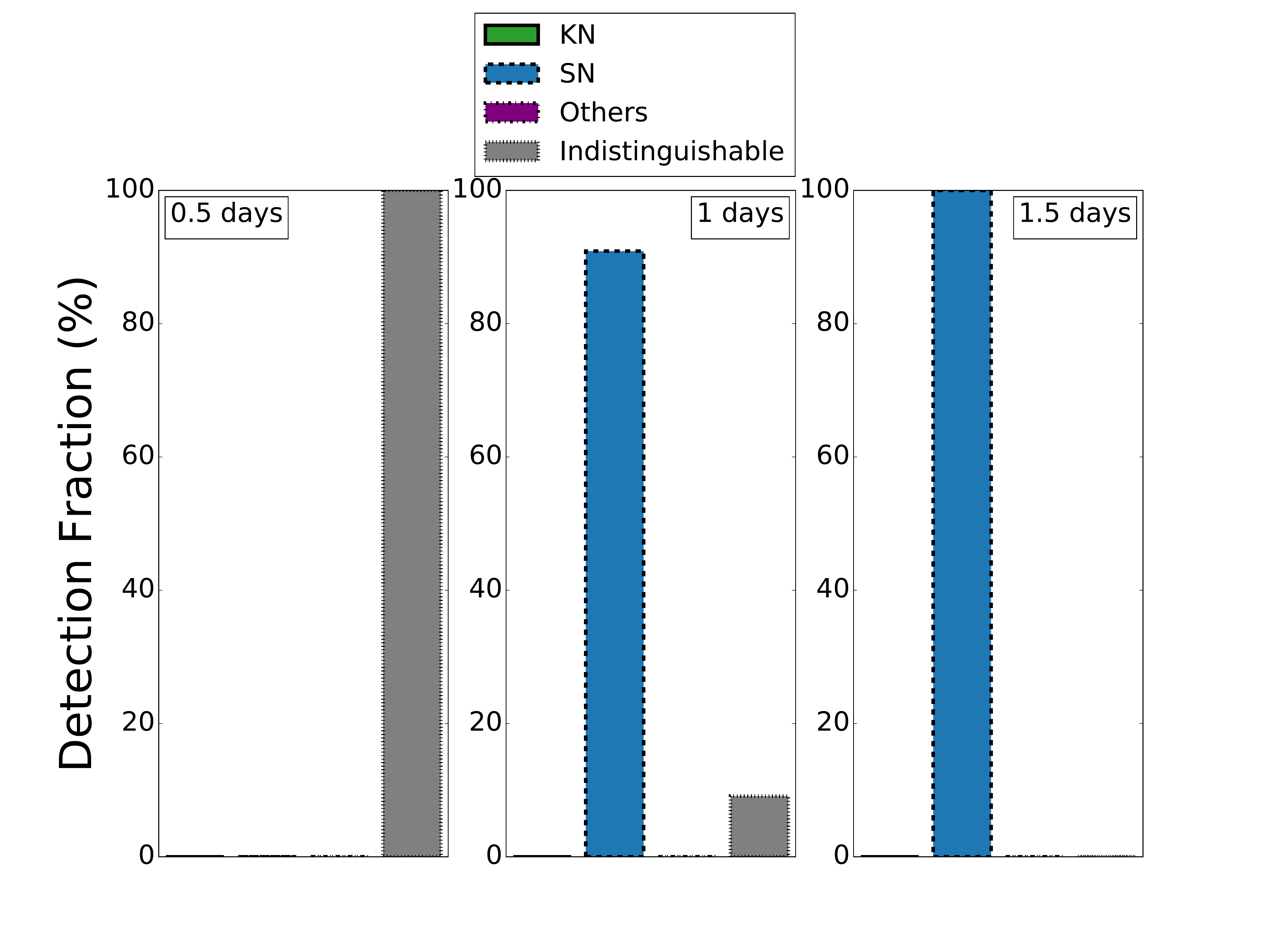}
    \caption{Preferred event fraction histogram  given different  amounts of observation time: from 0.5 day (left) up to 1.5 days (right). The input is represented by the set of 1000 SN injections.}
    \label{fig:SN-evolution}
\end{figure}

In this subsection, the lightcurves are ideally sampled (2 observations per night taken in each consecutive night) with detections in two filters ($r$-band and $g$-band).
In the case of the Type Ia supernova injections (see Figure~\ref{fig:SN-evolution}), the ``SN'' template is preferred after only a few nights. This is because these are the brightest transient events, and therefore their brightness is the key determinant.  

Also, given that these injections present only-rising, only-decreasing and rising-decreasing shapes, one can conclude that the classification efficiency is not simply due to shape recognition. The better classification efficiency with respect to the case of real ZTF ``SN'' type objects seems to arise from the sampling cadence. 

\subsubsection{Well-sampled kilonovae}
\begin{figure}[!htb]
    \centering
    \includegraphics[scale=0.18]{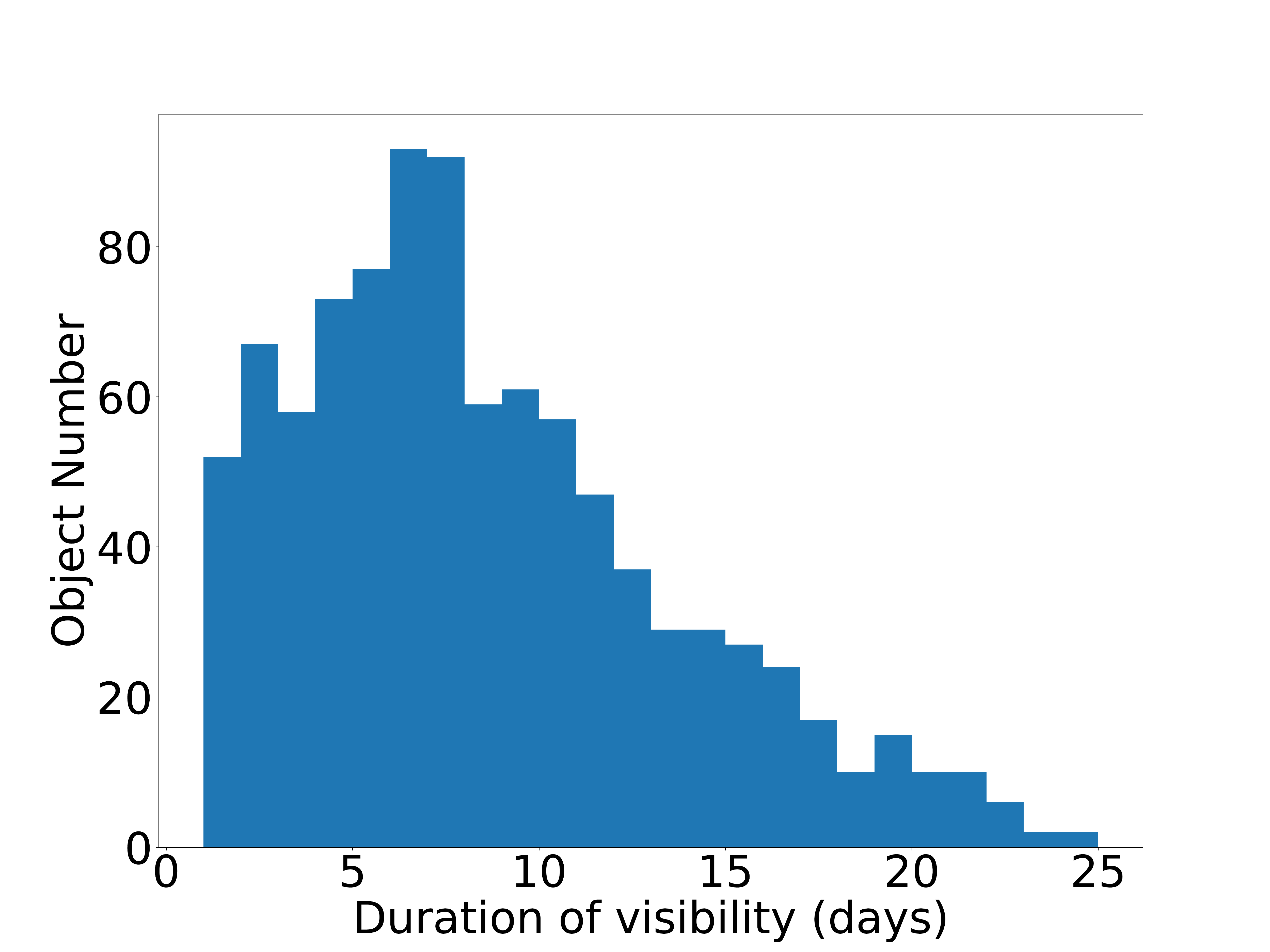}
    \caption{The number of days the kilonova is detectable by ZTF. We use the same set of 1000 ideally sampled KN injections.}
    \label{Fig:KN_days_obs}
\end{figure}

In this subsection, the kilonovae lightcurves considered are also ideally sampled (2 observations per night taken in each consecutive night) with detections in two filters ($r$-band and $g$-band). Given their inherent faintness, the kilonovae are not visible at the ZTF sensitivity at large distances as the supernovae. More precisely, the injection set of kilonovae are visible an average of 8.1 days. And the farther away a kilonova is, the less time it will be detectable by ZTF. A histogram showing the number of days the injected kilonovae are visible is in Figure~\ref{Fig:KN_days_obs}. 
As in the case of real ZTF objects, we consider a kilonova visible for only $N$ days, that the preferred event after $m$ days, with $m > N$, to be the same as the preferred event at the end of the real $N$ days of observation.

\begin{figure}[!htb]
    \centering
    \includegraphics[scale=0.18]{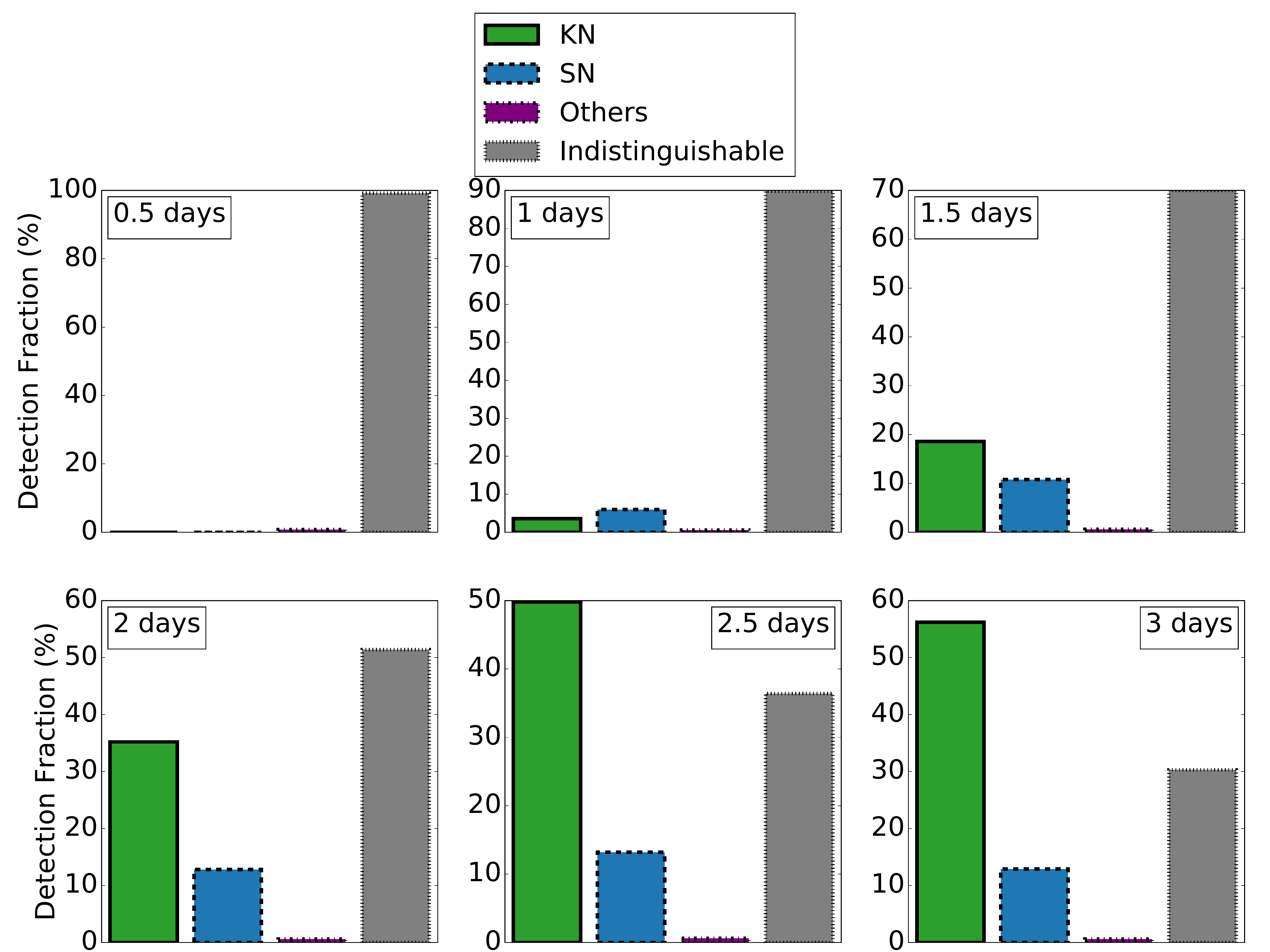}
    \caption{Preferred event fraction histogram  given different  amounts of observation time: from 0.5 days (top-left corner) up to 3 days (bottom-right corner). The input is represented by the set of 1000 kilonovae injections.}
    \label{Fig:KN-evolution}
\end{figure}

\begin{figure*}[!htb]
   \begin{minipage}{0.47\textwidth}
     \centering
     \includegraphics[width=1.1\linewidth]{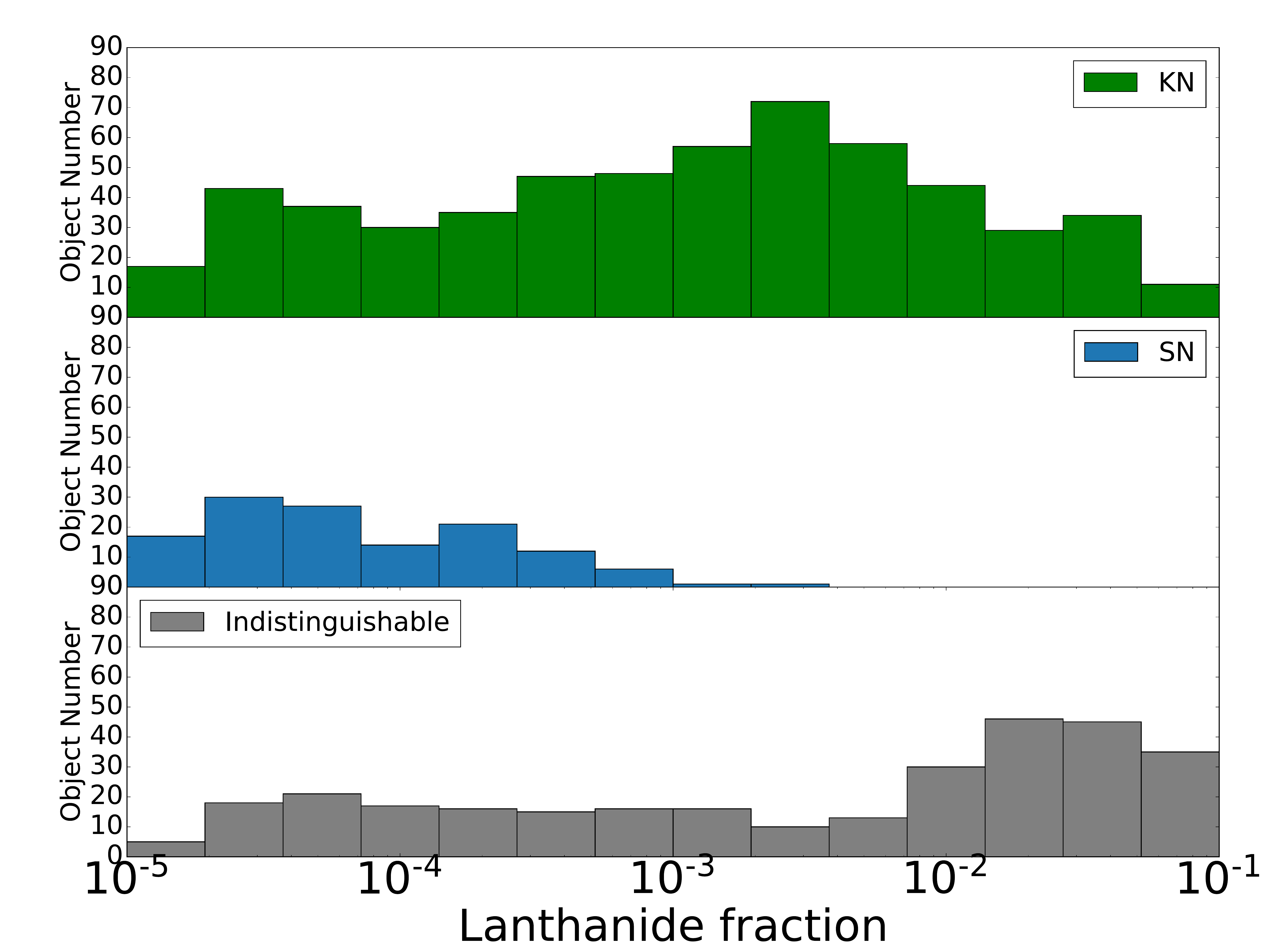}
   \end{minipage}\hfill
   \begin{minipage}{0.47\textwidth}
     \centering
     \includegraphics[width=1.1\linewidth]{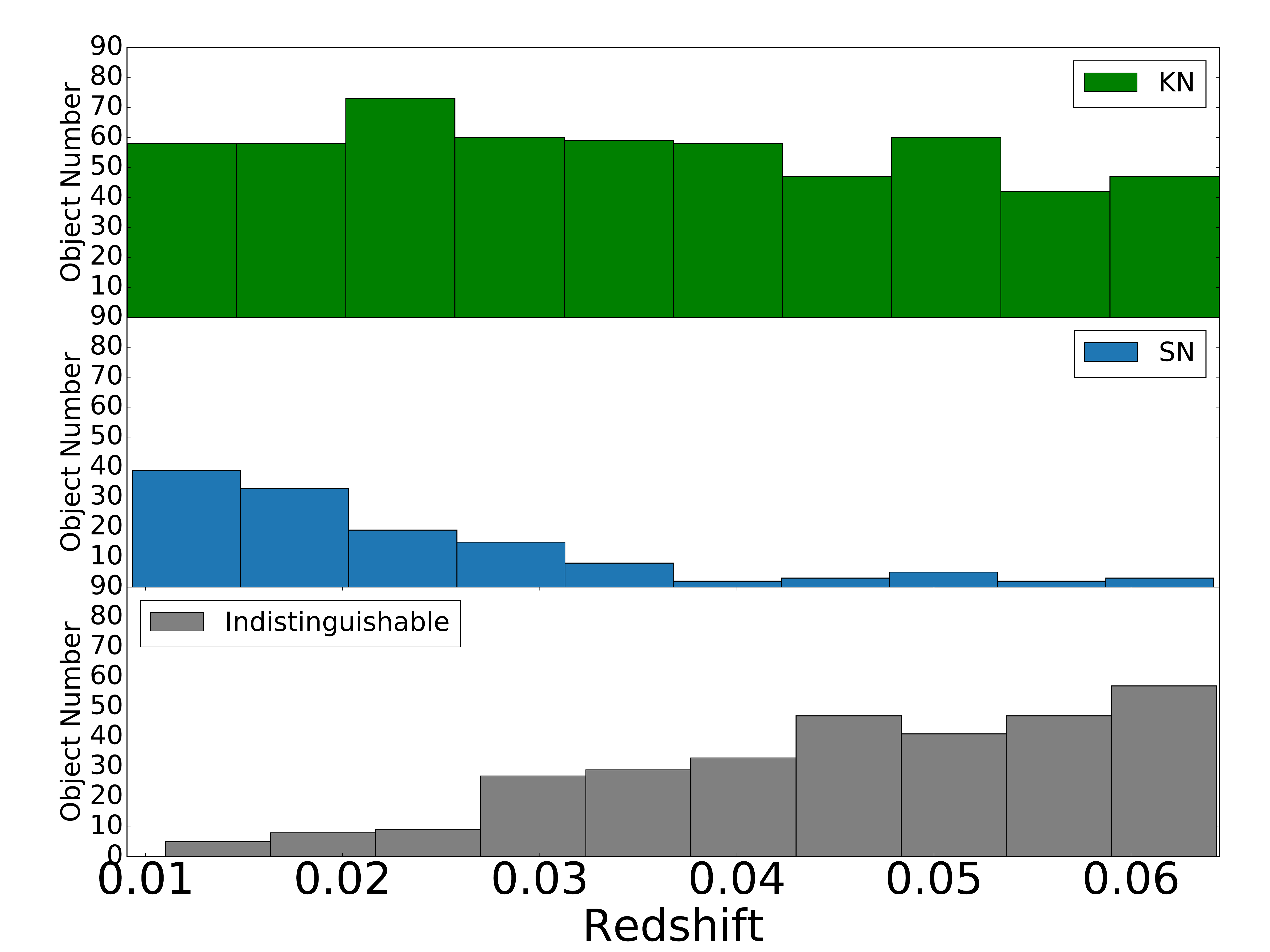}
   \end{minipage}
\begin{minipage}{0.47\textwidth}
     \centering
     \includegraphics[width=1.1\linewidth]{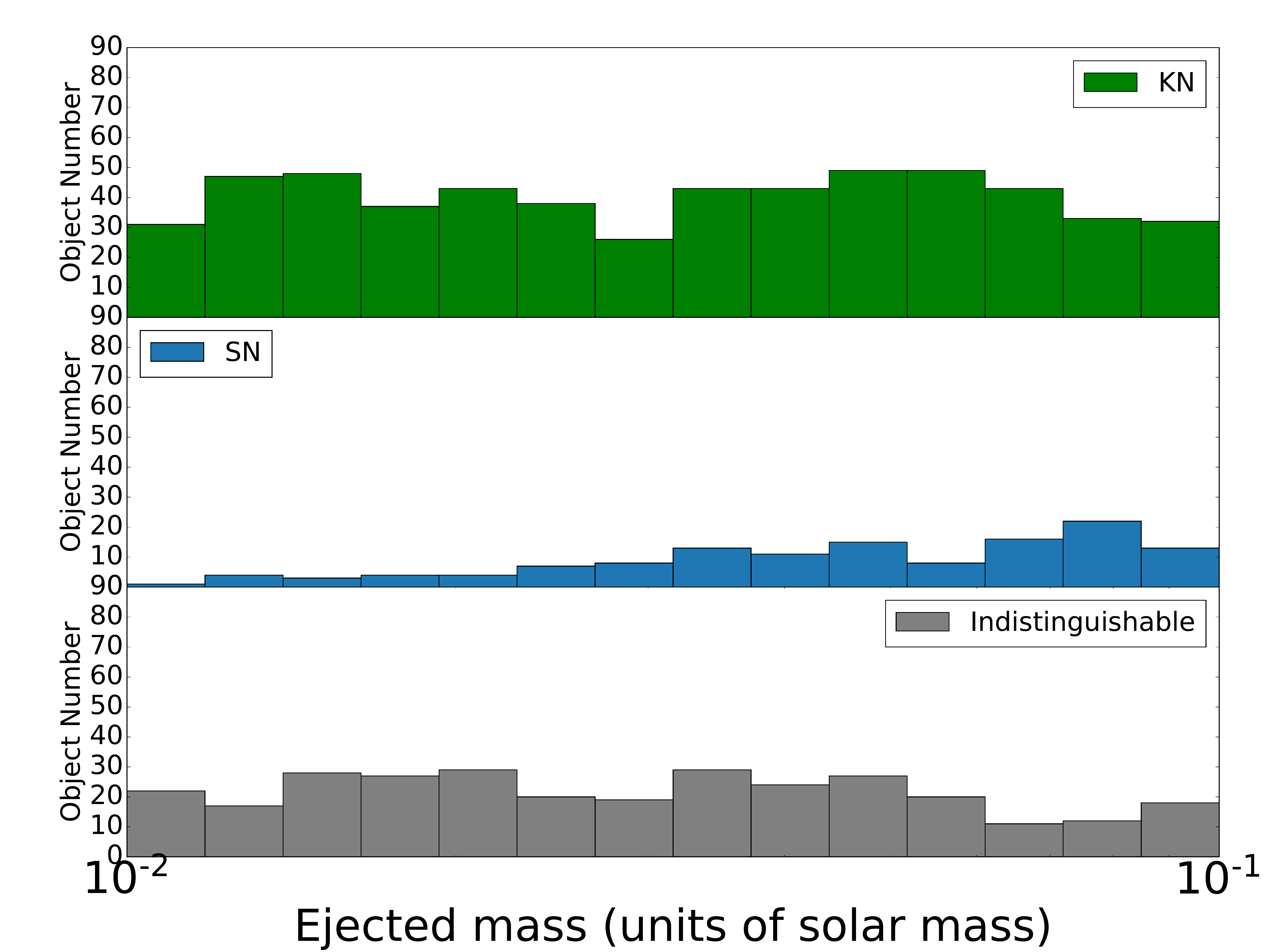}
   \end{minipage}\hfill
   \begin{minipage}{0.47\textwidth}
     \centering
     \includegraphics[width=1.1\linewidth]{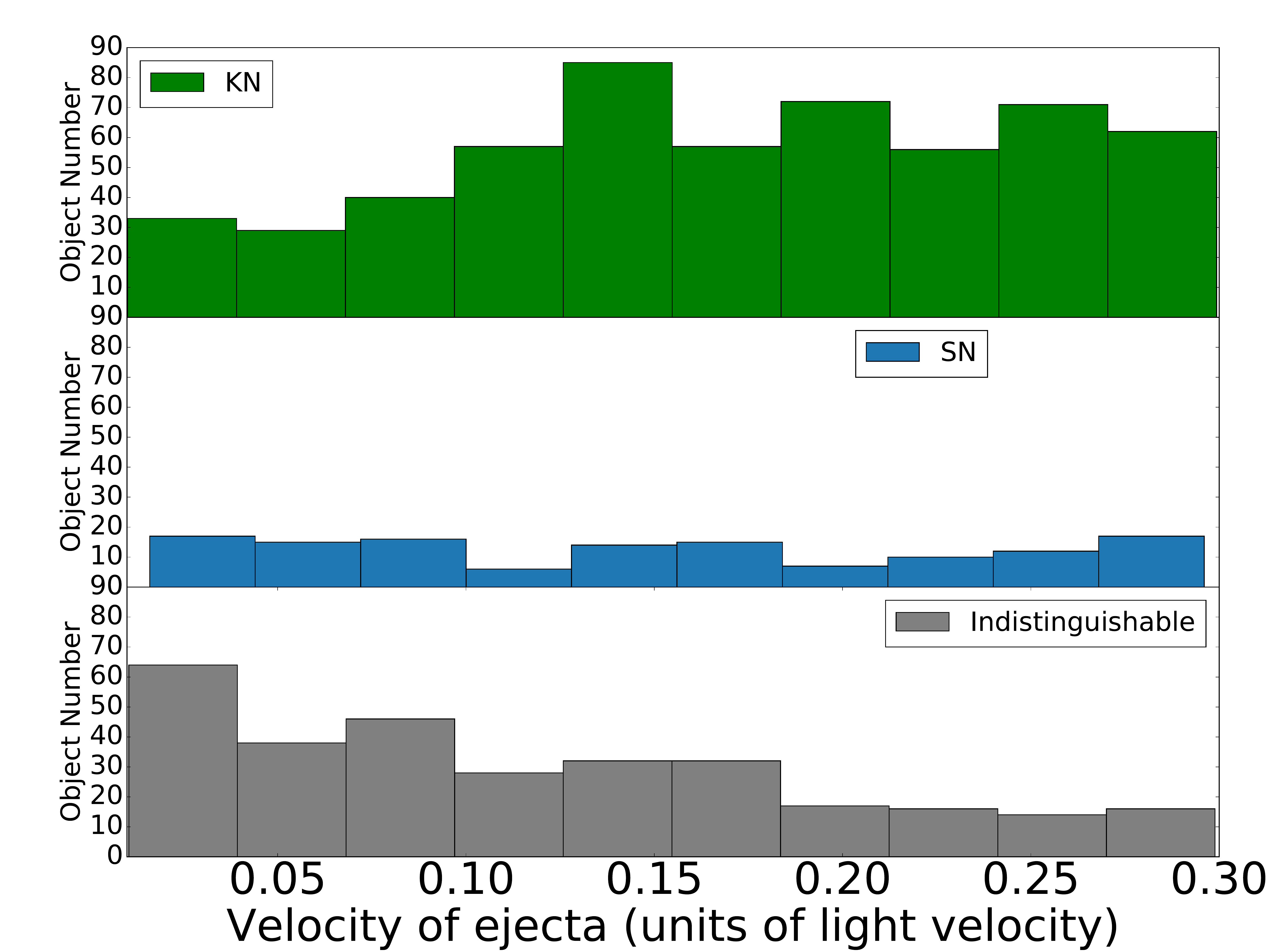}
   \end{minipage}
   \caption{Classification result dependence on the varied parameters for the 1000 KN injections. On the top row, a histograms of classifier favored events after 3 days of observation as a function of $X_{lan}$ (on the left) and redshift (on the right). On the bottom row is the same as a function of $M_{ej}$ (on the left) and $v_{ej}$ (on the right)}.
   \label{fig:KN-Xlan_redshift-mej_vej}
\end{figure*}

 Figure~\ref{Fig:KN-evolution} presents the histograms of preferred events for the very first night of observations. As expected, the category ``Indistinguishable'' is favored when there is not enough observations (usually less than $\sim$\,6), but at the end of two nights, the classifier identifies it as a kilonova lightcurve with a high probability. From Figure~\ref{Fig:KN-evolution}, two conclusions can be made. First, one can see that after only two days of observations, the ``KN'' template starts to be preferred over the others. Second, one notices that after 3 days of observations, the classifier mainly chooses the ``KN'' template, and the few failures consist of a preference for the ``SN'' and ``Indistinguishable'' templates. The false dismissal rate is less than 4 per 10 events. Figure~\ref{fig:KN-Xlan_redshift-mej_vej} shows the dependence of the classifier's preferred events (at the end of 3 days of observation) as a function of $X_{lan}$, redshift, $M_{ej}$ and $v_{ej}$.

Figure~\ref{fig:KN-Xlan_redshift-mej_vej} suggests that a low velocity for the ejecta has the effect of preferring ``Indistinguishable'' by the classifier. Likewise the more mass ejected leads to a higher preference for the ``SN'' templates; this is because they become brighter and therefore more consistent in appearance with supernovae. It also reveals that the lanthanide fraction and the redshift have a non-negligible impact on the lightcurve; we see that the smaller $X_{lan}$, the more probable is the ``SN'' template, and the higher $X_{lan}$, the more probable is ``Indistinguishable.'' The lanthanide fraction is responsible for the reddening of the lightcurve. As the lanthanide fraction decreases, the transient becomes more blue and therefore compatible with a supernova lightcurve (which are in general more gray). In addition, a large lanthanide fraction leads to red lightcurves, where the $g$-band observations are upperlimits instead of detections and so the ``Indistinguishable'' template becomes the favored one, as there is no color information. At small redshift, kilonovae can be misclassified as SNe (intrinsically brighter) at higher distance with respect to the injected one. On the other hand, a high value of the redshift denotes a small signal to noise ratio, and thus the classifier is also more likely to prefer the ``Indistinguishable'' template. 

\begin{figure}[!htb]
    \centering
    \includegraphics[scale=0.18]{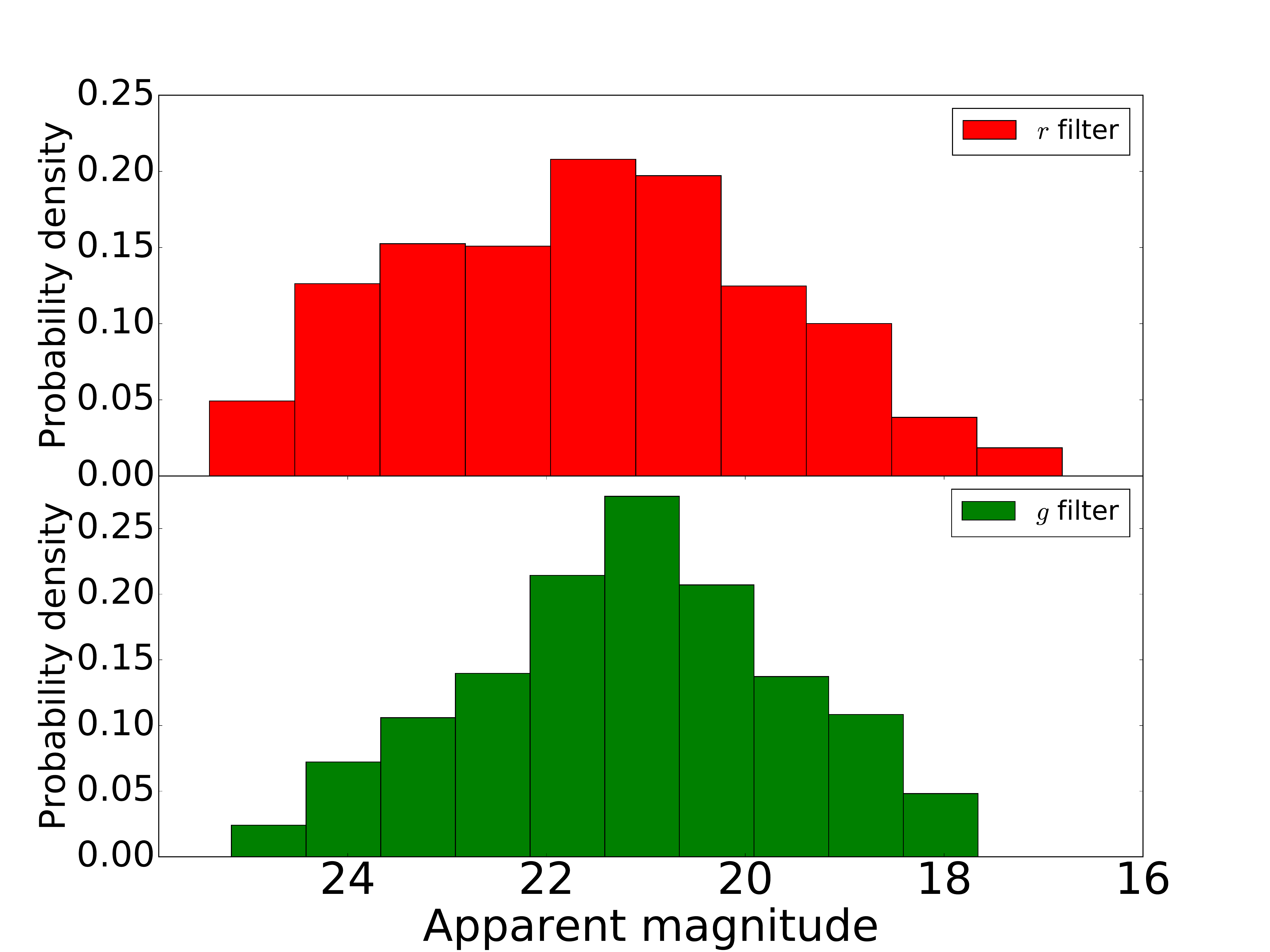}
    \caption{The probability density function for the apparent magnitude (on top for the $r$ filter, on bottom for the $g$ filter) at $t_1$, the first time the kilonova was identified as being "KN". It was considered a sample of 757 (553) lightcurves for $r$ ($g$) filter.}
    \label{fig:KN-brightness}
\end{figure}

In order to assess the utility of our method, understanding the brightness of the kilonovae once they are detected for the first time by the algorithm is of great significance. To this end, we pick out only those lightcurves which were classified as ``KN'' after the last observation. There are 757 such injections. For each of these lightcurves, we looked for the first time $t_1$ when our tool identifies it as a ``KN,'' and then we determine the magnitude of the kilonova at this particular time $t_1$. Whereas all of these lightcurves contain observations in the $r$ passband at $t_1$, only 553 of them are still detectable in the $g$ passband at $t_1$. This is because in general the fade in $g$ is faster than the fade in $r$ so usually that is the filter being measured last. In Figure~\ref{fig:KN-brightness}, we show the normalized histogram of the apparent magnitude with both filters at the time $t_1$. We can infer from this figure that in the majority of cases, our tool identifies the kilonovae early enough for follow-up by telescopes that can reach $\sim 21$\,mag, i.e. reasonably sensitive telescopes.  

\begin{figure}[!htb]
    \centering
    \includegraphics[scale=0.5]{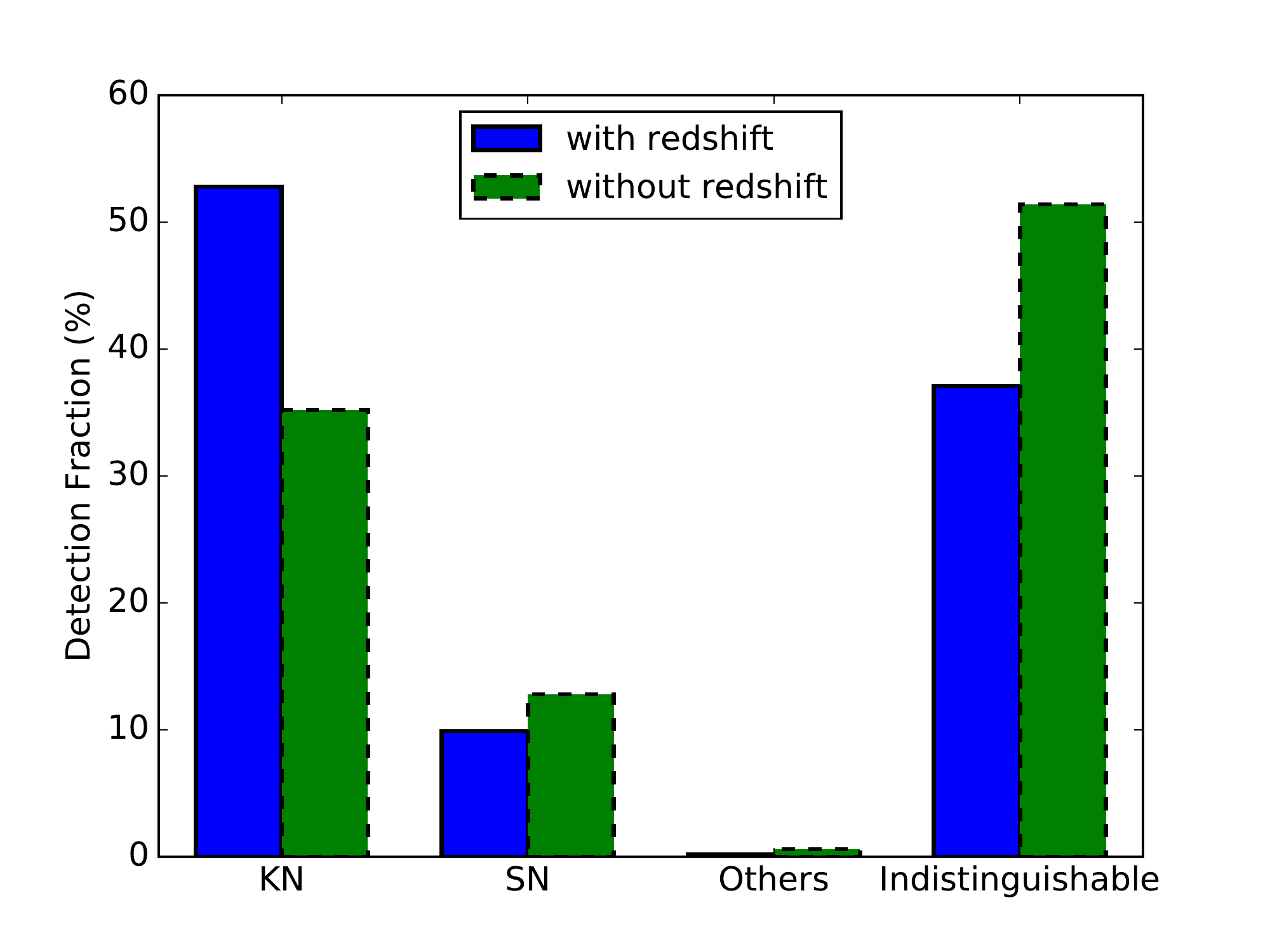}
    \caption{Classification results after 2 nights of observations with (blue) and without (green) the knowledge of the cosmological redshift. The input is represented by the same 1000 ``KN'' injection set discussed inside this subsection.}
    \label{fig:plus_minus_redshift}
\end{figure}

In the previous paragraphs, we presented the results of our method when the input consists only of photometric data. But as any classifier based on machine learning, \text{\astrorapid} provides more precise classifications if additional information is available. One example of such additional information might be the redshift of the candidate. In Figure \ref{fig:plus_minus_redshift} this is illustrated by means of a histogram for the quantitative improvement in the classification results when the redshift is taken into account compared to the situation where this information is considered unknown. For this study, we use the same set of well-sampled kilonova injections presented at the beginning of this subsection. One can observe that at the end of two nights of observations, the success rate in the recovery of KN templates is improved by more than 10\% if the cosmological redshift is given. 

\subsubsection{Realistically-sampled kilonovae}
While the case of dedicated ToOs was treated in the previous subsection by considering an ideal sampling of two observations per night and in each of the two filters $r$- and $g$-band, we are now interested in what is the efficiency of the classification in the case of a serendipitous observation. To this end, the spacing in time of two consecutive observations was studied for the case of real ZTF objects.
\begin{figure}[!htb]
    \centering
    \includegraphics[scale=0.18]{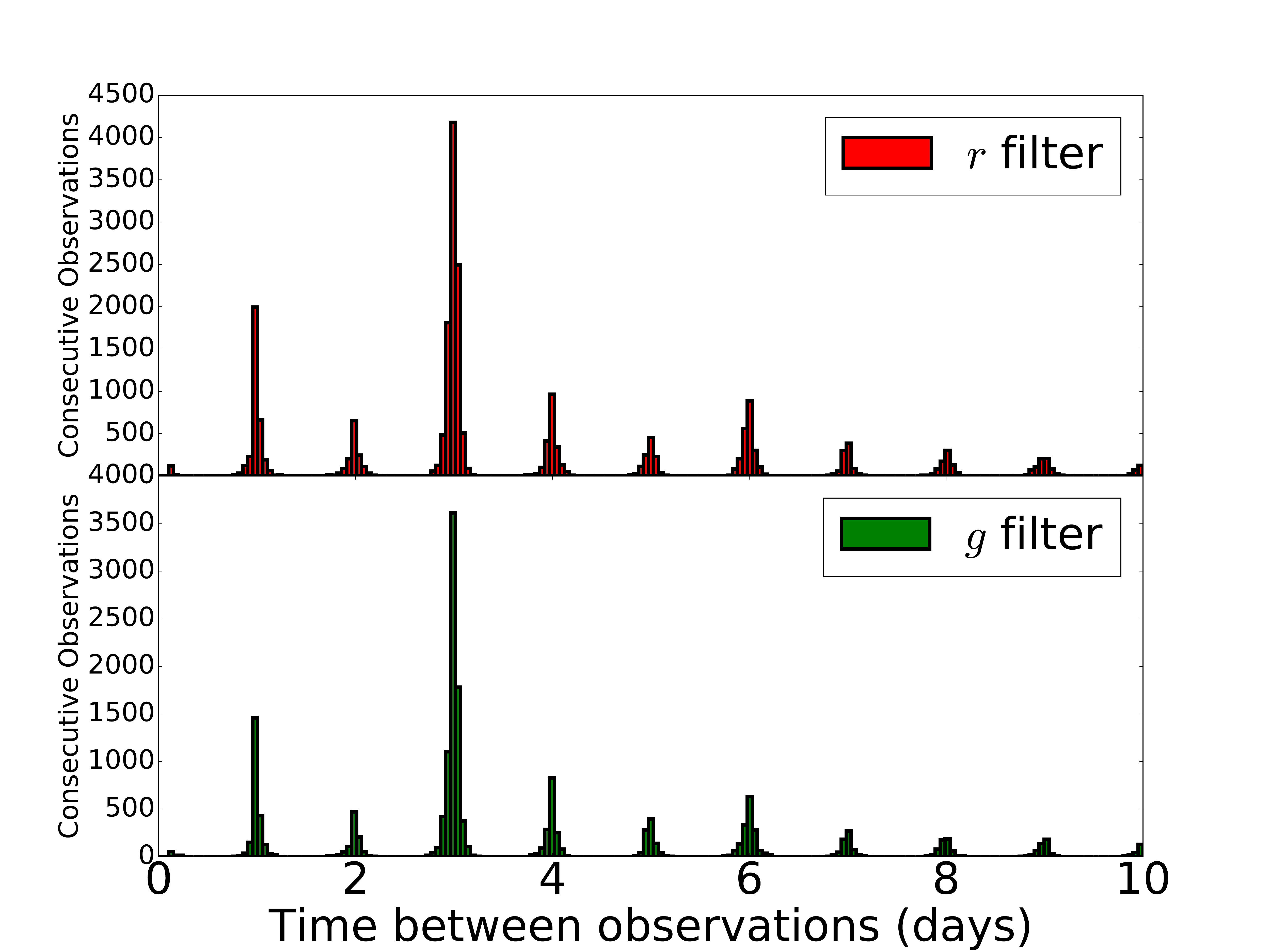}
    \caption{Histogram of time spacing between two consecutive observations in each filter. All real ZTF objects have been used for this study. The $r$-band results are on top and the $g$-band results on bottom.}
    \label{fig:ZTF_cadence}
\end{figure}
Indeed in Figure~\ref{fig:ZTF_cadence}, one can see the time difference between two consecutive observations in each filter for the entire set of real ZTF objects, both ``SN'' and ``Others'' types. One can easily observe a peak corresponding to a time spacing of 3 days. This is why we simulated a second set of 1000 kilonovae, with a sampling rate of 1 observation every 3 nights in each filter.
\begin{figure}[!htb]
    \centering
    \includegraphics[scale=0.18]{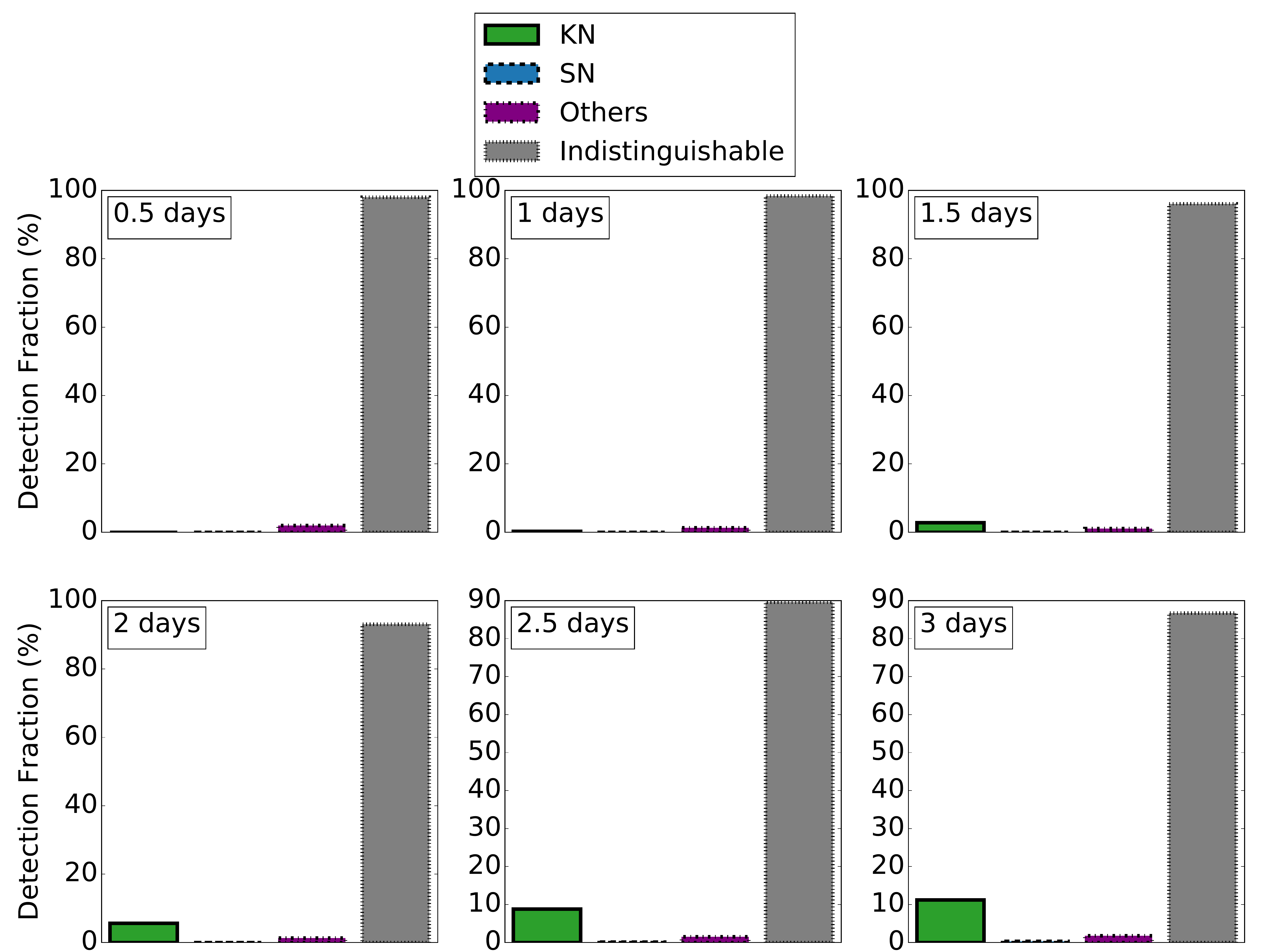}
    \caption{Preferred event fraction histogram  given different  amounts of observation time: from 0.5 days (top-left corner) up to 3 days (bottom-right corner). The input is represented by the set of 1000 kilonovae injections sampled similarly to the ZTF cadence.}
    \label{Fig:KN_cadence_evolution}
\end{figure}
Figure~\ref{Fig:KN_cadence_evolution} shows the results of the classification for the first 3 days of observations for this new set of injections. One can see that the classifications are much less efficient than in the case of the ideal sampling, which allows us to conclude that dedicated ToOs vs. serendipitous observations make a big difference.

\section{Conclusion}
\label{sec:conclusions}
In this study, we present a method to classify transients starting from photometric observations and the cosmological redshift. The method is based on  the use of an open-source classifier \text{\astrorapid}. By running this tool on input data represented by the observational lightcurves and combining the output results by source class, we propose a way to distinguish between four main classes: ``KN",'' ``SN",'' ``Others'' and ``Indistinguishable.'' The performance of this classifier and class system have been tested on both real ZTF objects and simulated lightcurves. The case of real transients from the public ZTF alert stream emphasizes the necessity of around 10 observations at the ZTF cadence,  provided that the information from both $r$ and $g$ passbands is used. Concerning the well-sampled lightcurves, it has been shown that the identification of SNe necessitates only a few observations, while for the recognition of KNe, a few nights of photometry in multiple passbands is required (with significantly worse results in the case of single passband observations). Finally, it has been shown that for kilonovae sampled at a cadence similar to that of ZTF, the efficiency of the classifier decreases significantly.

The current study opens prospects for future work, including the dominant question in the community concerning how to determine the best observing strategy to adopt in the case of a GW alert. In order to address this question, several issues should be considered. First of all, an evaluation of the candidate recognition dependence on the observing cadence needs to happen. In addition, a performance comparison between imaging in two filters and spending twice as long in a single filter needs to happen (with of course the possibility of even more filters being considered). Also, the possibility of having three or more filters cannot be discredited. Once the entire parameter space is explored, an optimal observing strategy could be found. The current version of \astrorapid \text{} does not have the current capability to address these items. To do so, we will need to retrain \astrorapid \text{} on single passband lightcurves, as well as lightcurves observed in more than three filters.

In the future, we intend to improve \astrorapid \text{} to account for ``missing'' observations where only upperlimits are available. This should be useful in cases of particularly red or blue transients, where likely the color information is even \emph{more} apparent than in the case where there are detections. Another improvement to \astrorapid \text{} might consist of the introduction in the training set of new  interlopers such as M-dwarf. In addition, we plan to incorporate these techniques into some of the ongoing follow-up infrastructure, such as the GROWTH target of opportunity marshal \citep{CoAh2019} and the GRANDMA iCARE pipeline \citep{Antier:2019pzz}.

\section{Acknowledgments}
MC is supported by the David and Ellen Lee Prize Postdoctoral Fellowship at the California Institute of Technology. The authors thank the Observatoire de la C\^{o}te d'Azur for support.

\appendix
\begin{appendices}

\section{Penalty factor and choice of threshold}
\label{sec:penalty factor}
When collapsing the probabilities of the fourteen initial \text{\astrorapid} templates, one could naively sum the corresponding probabilities, i.e.

\begin{eqnarray*}
& P(\text{SN}) & =  P(\text{SNIa-norm}) + P(\text{SNIbc}) + P(\text{SNII}) + P(\text{SNIa-91bg}) + P(\text{SNIa-x}) + P(\text{point-Ia}) + P(\text{SLSN-I}) + P(\text{PISN}) \\
& P(\text{Others}) & = P(\text{ILOT}) + P(\text{CART}) + P(\text{TDE}) + P(\text{AGN}) \\
& P(\text{KN}) & =  P(\text{Kilonova}) \\
& P(\text{Pre-explosion}) & =  P(\text{Pre-explosion})
\end{eqnarray*}

 Given the abundance of SN classes, ``SN'' would almost always be selected as the preferred class.

\begin{figure*}[!htb]
    \centering
    \includegraphics[scale=0.4]{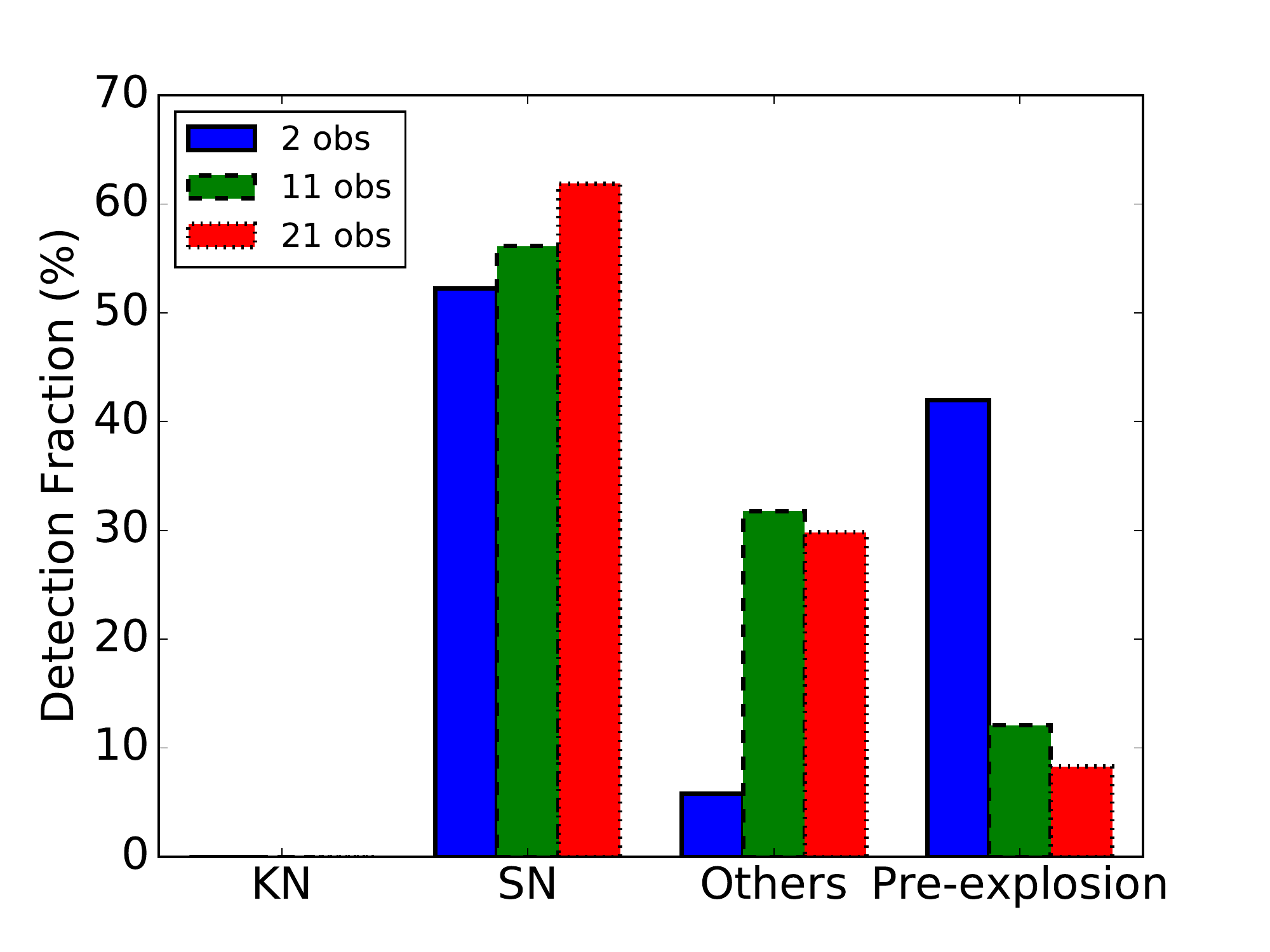}
    \includegraphics[scale=0.4]{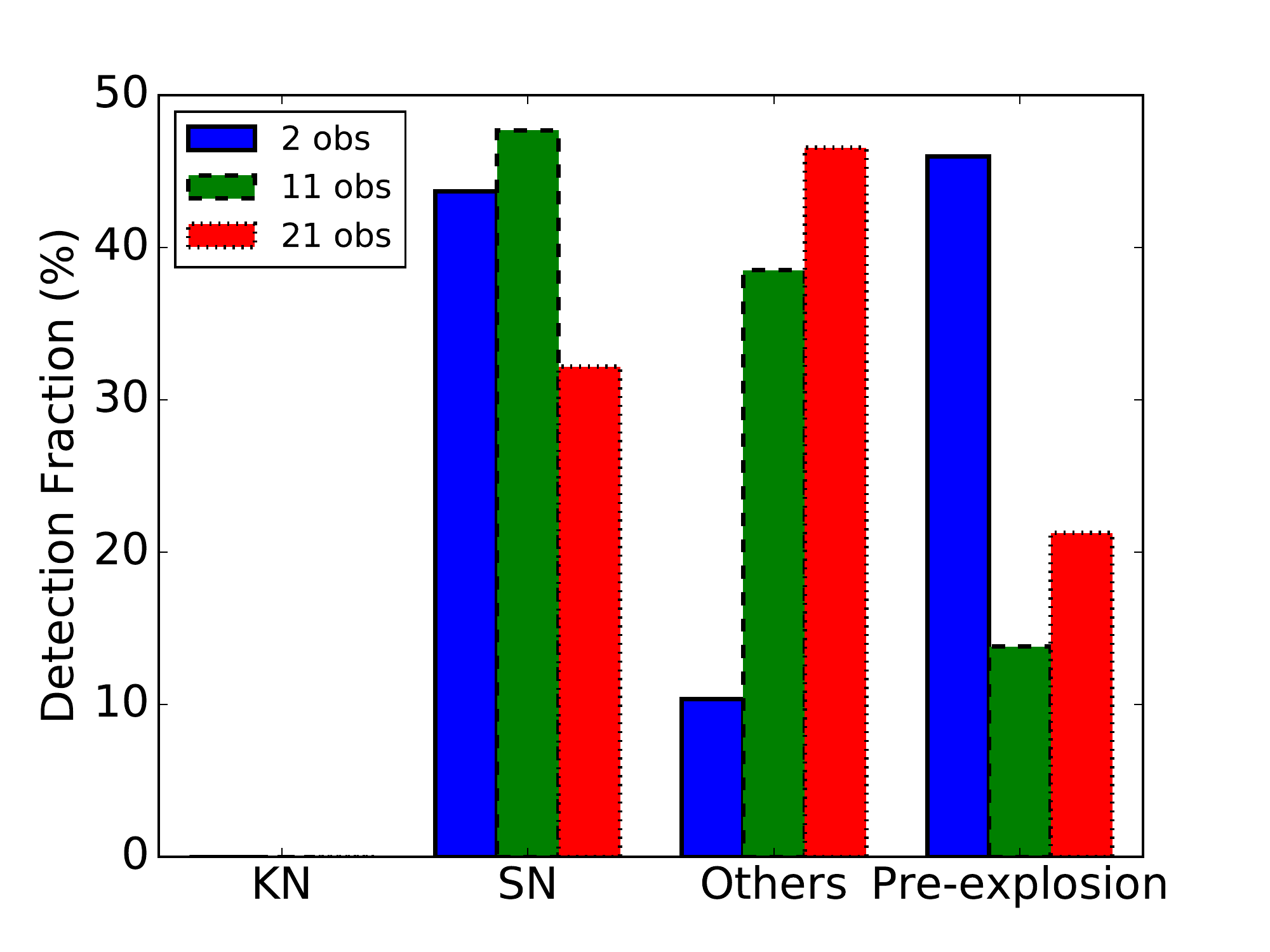}
    \caption{The preferred event after 2, 11 and respectively 21 observations. The input is represented on the left by the set of 2,049 ZTF real sources identified as ``SN'' type and on the right by the set of 174 ZTF real sources identified as the ``Others'' type.}
    \label{fig:preferred-vierge}
\end{figure*}

We illustrate this behaviour on real ZTF lightcurves, i.e. the same data set we used in Section~\ref{subsec: real transients}. In Figure~\ref{fig:preferred-vierge}, we plot the detection fraction of the preferred event after different numbers of observations. From this figure, several conclusions can be drawn. First of all, one can see, especially in the case where ``Others'' lightcurves are input, the ``SN'' class is often incorrectly chosen. This behaviour is visible for early observations especially and improves somewhat at later times. This motivates our decision to penalize the ``SN'' class by a factor $(1.- e^{- k^{\text{th}}_{\text{obs}} / \beta})$, where $k^{\text{th}}_{\text{obs}}$ stands for the $(k+1)^{\text{th}}$ observation. Secondly, at early times, too many events are misclassified as ``Others'' and ``SN'', when a non-determinative choice, i.e. a preference for the ``Pre-explosion'' template is physically reasonable. This supports our decision to replace ``Pre-explosion'' with ``Indistinguishable'' as well as the introduction of a probability threshold in order for a transient to be classified in the set \{"Indistinguishable", "SN" and "Others"\} 

\begin{figure}[!htb]
    \centering
    \includegraphics[scale=0.25]{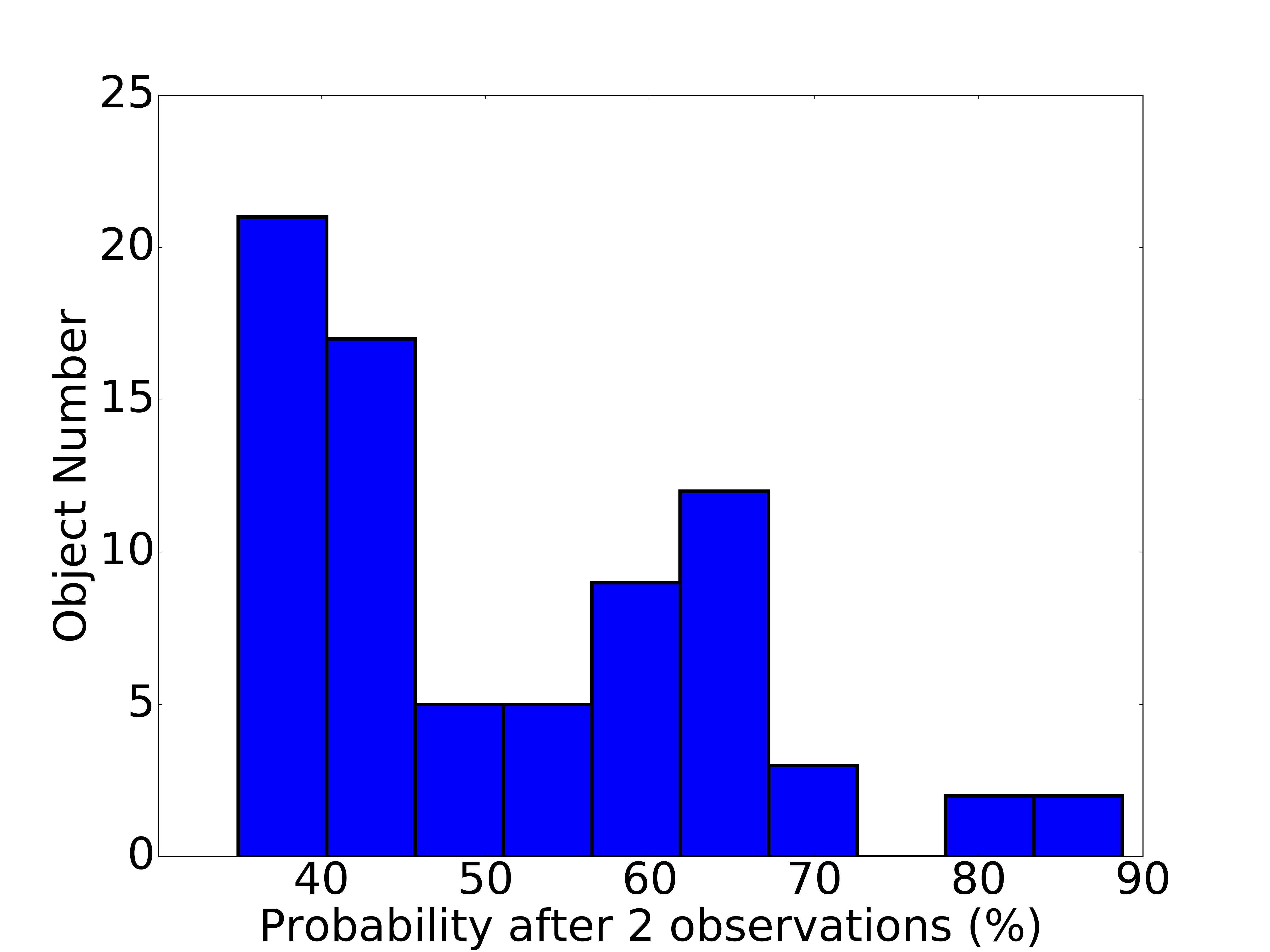}
    \caption{Histogram of the preferred event probabilities after the 2 observations. The input is represented by those real "Others" lightcurves which are misclassified as ``SN'' or ``KN'' after two observations.}
    \label{Fig:failure_prob}
\end{figure}
In Figure~\ref{Fig:failure_prob}, we illustrate the distribution of \text{\astrorapid} preferred event probabilities after 2 observations, when the input is a real ZTF ``Others'' type object, and the preferred event at the end of 2 observations is in \{"KN", "SN"\}. Figure~\ref{Fig:failure_prob}, therefore, shows the probability of the ``wrong'' preferred class when there is misclassification at early times. As a consequence of these results, we choose a threshold equal to 40\% below which the preferred event is always ``Indistinguishable''. 

Once the threshold is fixed, we are looking for a value of $\beta$ based on our data set results. Given the expression of the penalty factor we imposed, $(1.- e^{- k^{\text{th}}_{\text{obs}} / \beta})$, it is worth mentioning the $\beta$-dependence of our classifications. A too small $\beta$ means a penalty factor close to 1, so basically the classifications will be very similar to our initial results. On the other hand, a too large $\beta$ means a penalty factor close to 0, which will have as a consequence the preference of the other classes than ``SN.'' We can anticipate from this discussion that a near-optimal $\beta$ will be obtained as a tradeoff between these two regimes. To assess this, we analyze the impact of the penalty factor on the same set of real ZTF objects as presented in Section~\ref{subsec: real transients}. For the ``SN'' type objects, we define the failure probability as being the proportion of ``SN'' type objects classified as ``Others'' or ``KN''. Analogously for the ``Others'' type objects, we define the failure probability as the proportion of ``Others'' type objects classified as ``SN'' or ``KN''. It should be clear that these success/failure probabilities are function of the number of observations.

\begin{figure*}[!htb]
   \begin{minipage}{0.47\textwidth}
     \centering
     \includegraphics[width=1.1\linewidth]{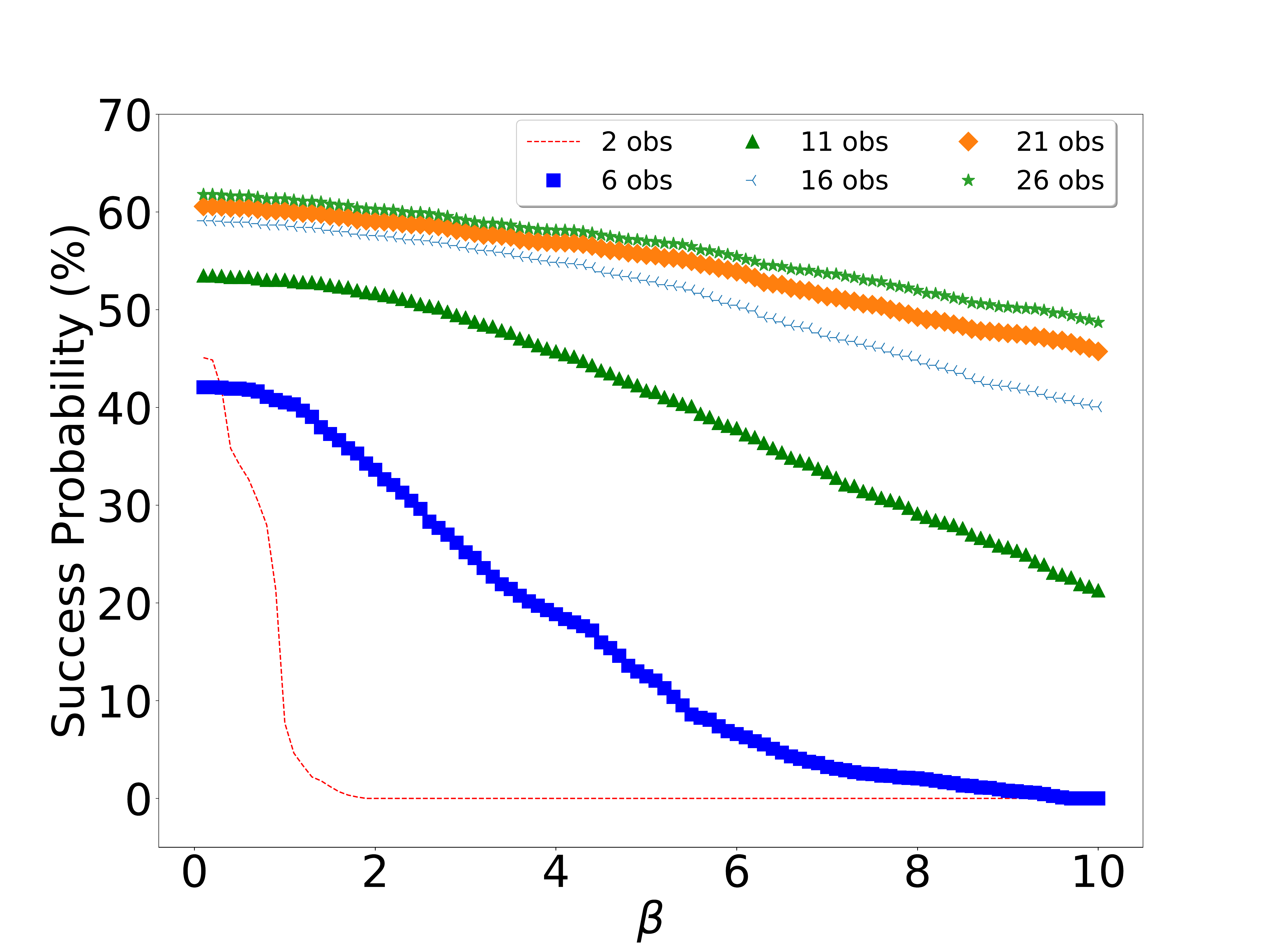}
   \end{minipage}\hfill
   \begin{minipage}{0.47\textwidth}
     \centering
     \includegraphics[width=1.1\linewidth]{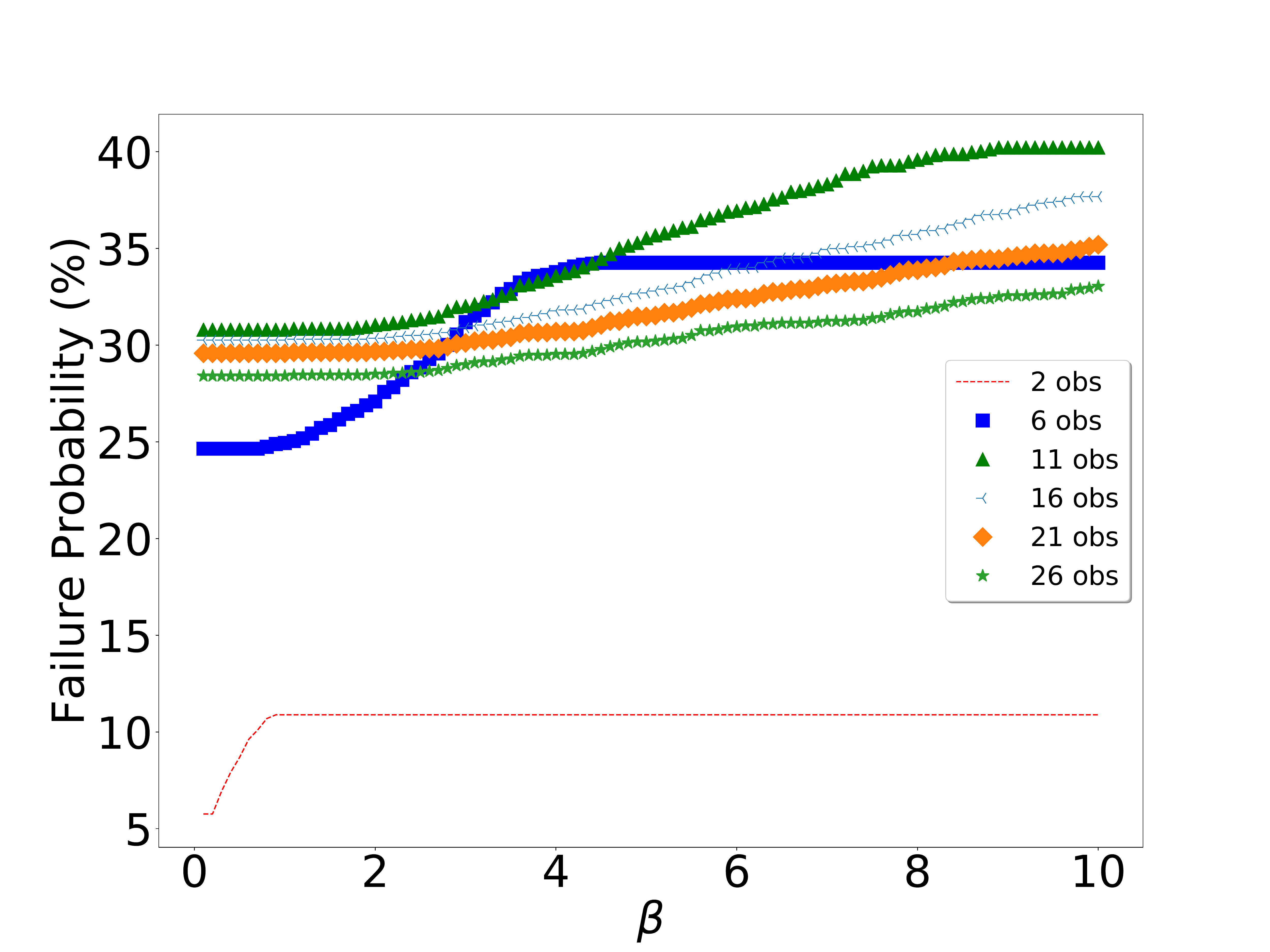}
   \end{minipage}
\begin{minipage}{0.47\textwidth}
     \centering
     \includegraphics[width=1.1\linewidth]{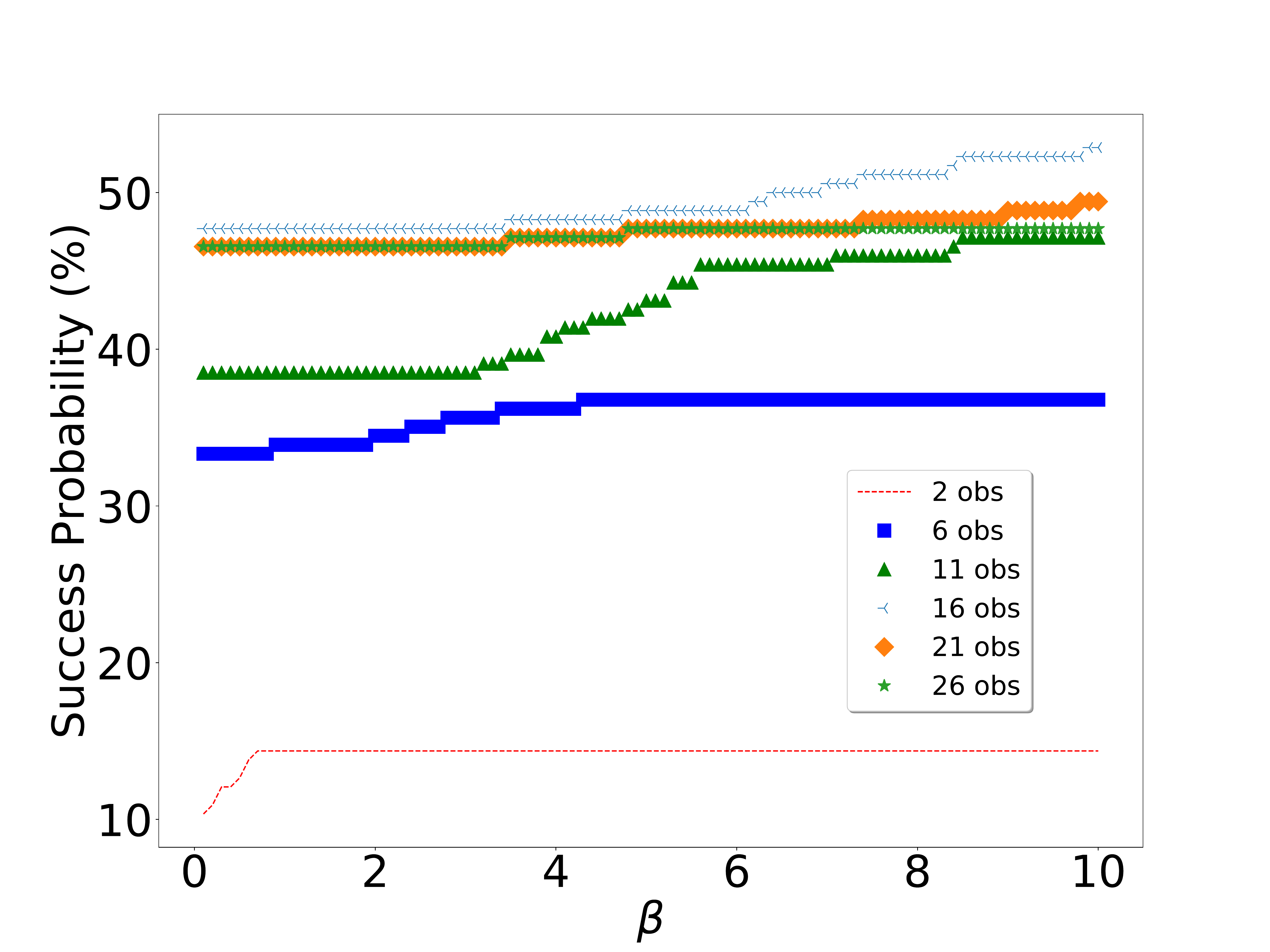}
   \end{minipage}\hfill
   \begin{minipage}{0.47\textwidth}
     \centering
     \includegraphics[width=1.1\linewidth]{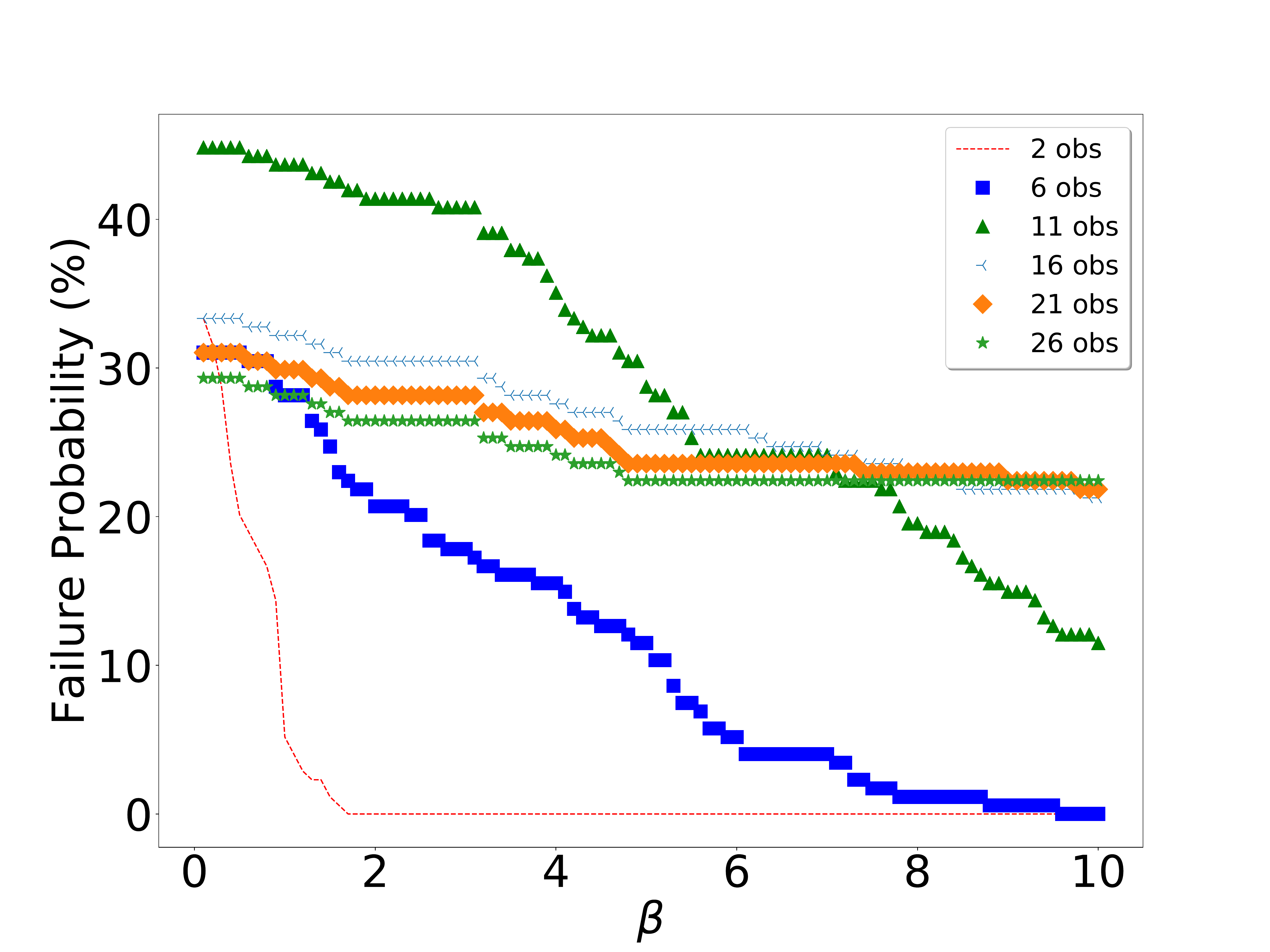}
   \end{minipage}
   \caption{Success and Failure probability for ``SN'' type objects (on top) and ``Others'' type objects (on bottom). It turns out that from these plots that $\beta = 4$ is a reasonable choice.}
   \label{fig:success_failure}
\end{figure*}

In Figure~\ref{fig:success_failure}, one can see that except for the very earliest times ($\sim$\,2 observations), all the other curves contain $\beta = 4$, indicating it is near optimal. 

\end{appendices}

\bibliographystyle{aasjournal}
\bibliography{references}

\end{document}